# Design and Optimization of Spin Dynamics in Ge Quantum Dots: g-Factor Modulation, Geometry-Induced Dephasing Sweet Spots, and Phonon-Induced Relaxation

Ngoc Duong,[1] and Daryoosh Vashaee[1,2,*]

[1]Electrical and Computer Engineering Department, North Carolina State University, Raleigh, North Carolina 27616, USA

[2]Materials Science and Engineering Department, North Carolina State University, Raleigh, North Carolina

## Abstract

Gate geometry and bias asymmetry can be used to engineer spin dynamics in gate-defined Ge hole quantum dots by reshaping the confinement potential and driving transitions between distinct confinement regimes. Using a realistic three-dimensional device model, we show that these transitions strongly modify wavefunction localization, heavy-hole/light-hole mixing, and the effective vertical electric field, leading to pronounced g-factor modulation and geometry-induced dephasing sweet spots where the qubit becomes first-order insensitive to vertical electric-field fluctuations. We further find that phonon-induced spin relaxation exhibits a strong dependence on device size and bias, with $T_1$ following a magnetic-field scaling close to $B^{-9}$, consistent with Rashba-dominated heavy-hole spin dynamics. Our results establish gate pattern and bias design as practical tools for optimizing spin coherence in Ge hole-spin qubits.

## I. Introduction

Previous implementations of spin qubits, such as electron spins in GaAs/AlGaAs quantum dots, suffer from limited coherence times due to hyperfine interactions with nuclear spins from Ga and As isotopes. Silicon-based electron spin qubits mitigate this issue by utilizing isotopically purified materials with negligible nuclear spin, but they face challenges associated with the valley degree of freedom. In particular, the valley splitting in Si quantum wells depends sensitively on atomic-scale disorder and interface roughness, making it difficult to control reproducibly and thereby complicating qubit initialization and manipulation.[1]

Hole spin qubits offer a compelling alternative. In particular, gate-defined hole spins in strained Ge quantum wells combine several attractive features, including the absence of valley degeneracy, compatibility with standard CMOS processing, and suppression of the contact hyperfine interaction due to the p-type orbital character of valence-band states. These advantages have led to growing interest in Ge-based hole spin qubits as a promising platform for scalable semiconductor quantum information processing. At the same time, the strong intrinsic spin-orbit coupling (SOC) of valence-band holes, arising from heavy-hole (HH) and light-hole (LH) mixing, introduces a competing challenge: enhanced sensitivity to charge noise, which can accelerate both spin relaxation and dephasing relative to electron-spin platforms.[2] In realistic quantum-dot heterostructures, such charge noise can originate from electrically active trap states at oxide interfaces, heterostructure boundaries, and within the semiconductor bulk; the resulting electrostatic fluctuations perturb the confinement potential and can couple to the qubit through spin-orbit interactions, contributing to dephasing and gate errors.[3]

* Corresponding author: dvashae@ ncsu.edu

This strong SOC also creates an important opportunity. Because hole spins couple efficiently to electric fields, they can be manipulated entirely electrically through electric dipole spin resonance (EDSR), without requiring local magnetic field gradients or oscillating magnetic fields. As a result, hole spin qubits can support Rabi frequencies in the 100 MHz to GHz range, enabling nanosecond-scale gate operations. Although coherence times in hole systems are often shorter than in electron-spin qubits, the substantially faster gate speeds can compensate, yielding comparable or even superior qubit quality factors, $Q = T_2^{\mathrm{echo}}/t_{\mathrm{gate}}$. For example, while an electron spin qubit may exhibit $t_{\mathrm{gate}} \sim 100$ns and $T_2^{\mathrm{echo}} \approx 100\ \mu$s, corresponding to $Q \sim 10^3$, hole spin qubits with $t_{\mathrm{gate}} \sim 1$ ns and $T_2^{\mathrm{echo}} \sim 1$–$10\ \mu$s can in principle reach $Q \sim 10^3$–$10^4$.

A central consequence of strong SOC in Ge hole systems is that both spin relaxation ($T_1$) and dephasing ($T_2^*, T_2$) become strongly anisotropic and highly sensitive to confinement, strain, and electric field. This can produce orientation-dependent operating points, including sweet spots where decoherence is suppressed. At the same time, it means that qubit properties are determined not only by the host material, but also by the detailed device geometry and gate-defined electrostatic environment. In realistic gate-defined quantum dots, the confinement potential is rarely well represented by an idealized parabolic or rectangular form. Instead, it is shaped by the full three-dimensional gate layout, heterostructure profile, and bias configuration, all of which can modify wavefunction localization, HH-LH mixing, effective electric fields, and ultimately the g-tensor and coherence properties of the qubit. This makes device-specific modeling essential for identifying experimentally relevant operating regimes and design rules.

Our broader research has emphasized that realistic modeling and coherence engineering must be considered together. In prior work, we proposed microwave-induced cooling schemes based on double quantum dots to reduce thermal noise near spin qubits and to engineer lower effective temperatures in coupled quantum systems.[4,5] More recently, we showed that reservoir-engineered refrigeration with double-quantum-dot spin qubits can be used to cool superconducting microwave cavities, and that quantum correlations can further enhance cavity cooling below the ambient temperature.[6] Complementing these environmental-control approaches, we have also shown that dynamic susceptibility and quantum Fisher information provide useful tools for characterizing entanglement and correlations in confined quantum-dot systems under strong interaction and spin-orbit regimes.[7] Together, these studies motivate the need for high-fidelity modeling frameworks that connect realistic device structure to coherence, control, and noise sensitivity.

While recent experiments have demonstrated sweet-spot operation in Ge hole spin qubits through electric-field tuning of a highly anisotropic g-tensor,[8,9] recent modeling has shown that sweet spots obtained by bias tuning alone can, in realistic planar Ge systems, be shifted to experimentally impractical electric fields for certain magnetic-field orientations.[10] This highlights the need for additional design knobs beyond electrostatic biasing alone. Device geometry is one such knob. As we show in this work, realistic three-dimensional gate-defined electrostatics can drive transitions between distinct confinement regimes, enabling geometry-induced modulation of the g-tensor and providing a route to shift or broaden sweet-spot operation through device design rather than bias tuning alone. Moreover, unlike prior studies that focus on individual aspects of hole-spin physics, we analyze confinement, HH–LH mixing, g-tensor anisotropy, Rashba-driven spin response, phonon-induced relaxation ($T_1$), and dephasing sensitivity ($T_2^*$) within a single realistic three-dimensional device model. This unified framework reveals how device geometry and electrostatics simultaneously influence multiple qubit performance metrics, and exposes design trade-offs and operating regimes that are not accessible in simplified or decoupled models.

Previous realistic-device calculations have established that inhomogeneous and nonseparable gate fields can generate additional g-tensor-modulation mechanisms and strongly affect electrical spin driving.[11] The present study addresses a complementary question: how the same type of realistic three-dimensional electrostatics can reorganize the static confinement regime and thereby correlate wavefunction localization, HH-LH mixing, g-tensor response, dephasing sensitivity, and phonon-induced relaxation within a single device model.

In this work, we employ a finite-element $\mathbf{k}\cdot\mathbf{p}$ simulation framework (QTCAD) to investigate how the planar dimensions, gate geometry, and bias asymmetry of $Si_{0.2}Ge_{0.8}$/Ge quantum dot devices influence the spin dynamics of confined hole states. Rather than treating geometry as a secondary perturbation, we show that the realistic three-dimensional gate-defined potential can drive transitions between distinct confinement regimes, accompanied by substantial changes in wavefunction position, extent, and HH-LH composition. These changes, in turn, produce strong modulation of the anisotropic g-tensor, enable geometry-induced dephasing sweet spots, and alter the phonon-mediated relaxation dynamics. By resolving these coupled electrostatic and spin-dependent effects within a realistic device architecture, this work provides concrete physical insight into how gate layout and bias design can be used as practical tuning parameters for optimizing coherence in Ge hole spin qubits.

## II. Methodology

### II.I Device

The simulated spin qubit device is based on a strained Ge quantum well embedded in a $Si_{0.2}Ge_{0.8}$/Ge/$Si_{0.2}Ge_{0.8}$ heterostructure. The full vertical stack consists of a 30 nm $Al_2O_3$ gate oxide, a 55 nm $Si_{0.2}Ge_{0.8}$ top barrier, a 16 nm intrinsic Ge quantum well, another 55 nm $Si_{0.2}Ge_{0.8}$ bottom barrier, and a 50 nm Si substrate. This composition creates a quasi-square potential well along the growth (z) direction, confining holes within the Ge quantum well and supporting the formation of a two-dimensional hole gas.

We assumed a strain-relaxed $Si_{0.2}Ge_{0.8}$ barrier layer, and the residual strain in Ge well layers is calculated by assuming a uniform lattice constant throughout the heterostructure stacks. As a result, the strain in Ge layer solely depends on the Si content in the barrier layer.

Lateral confinement is achieved via a depletion-mode gate design, in which metal gates are patterned directly on the $Al_2O_3$ dielectric layer. Positive bias on the gates locally depletes the hole gas, forming quantum dots through electrostatic confinement in the in-plane directions. This architecture allows for tunable confinement potential via gate voltage control. The following gate design offers a stable potential landscape under the chosen gate voltage control range and dimension changes, making it more suitable for spin-dependent calculations.

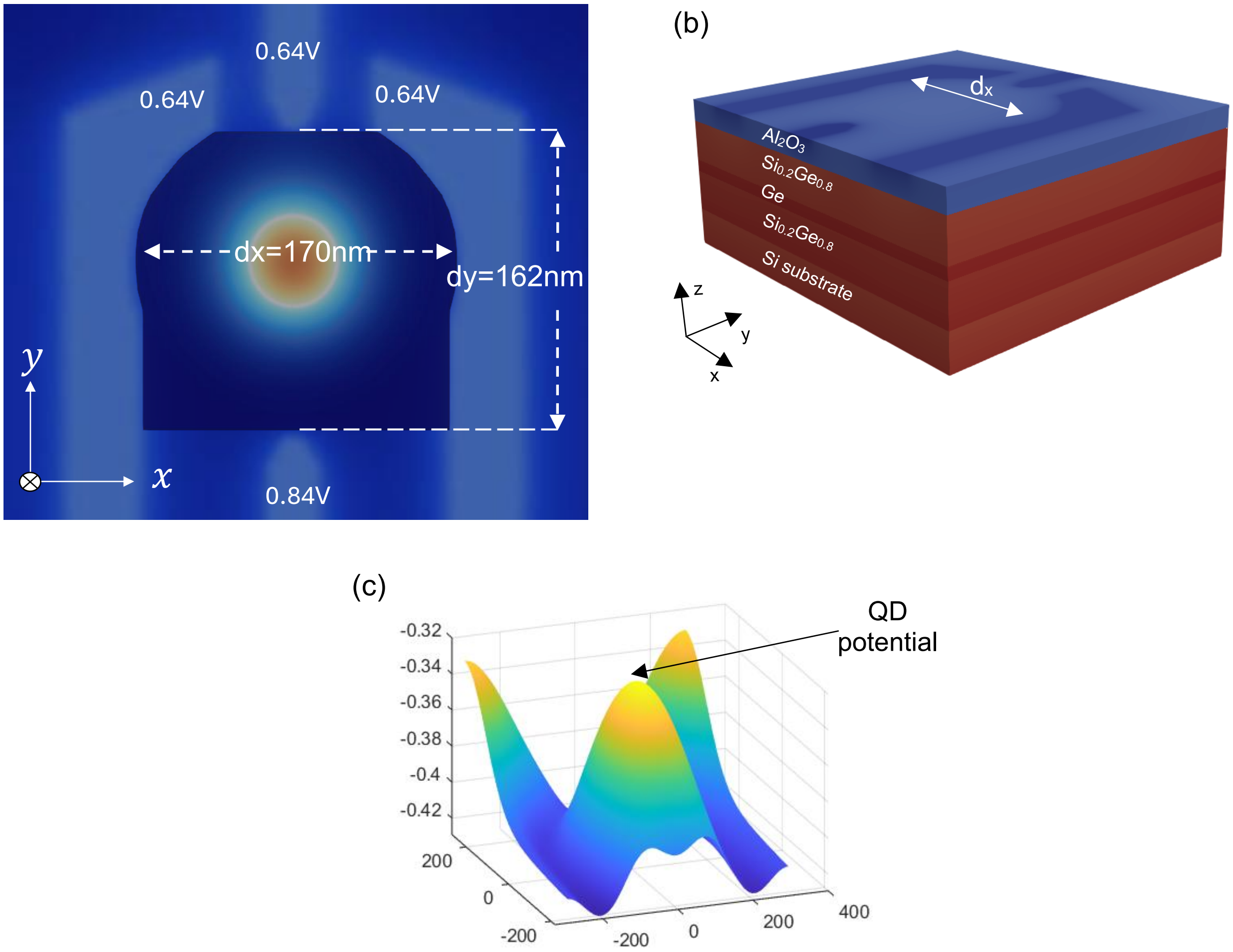

Figure 1: Electrostatic potential energy distribution for the $Si_{0.2}Ge_{0.8}$/Ge heterostructure under different gate configurations, obtained from self-consistent Poisson-Schrödinger simulations. (a) Surface potential at the $Al_2O_3$ interface for the gate layout. (b) Surface potential on a 3D view of the simulated device geometry. (c) In-plane potential profile (xy plane) within the Ge quantum well corresponding to Size 1. In (a), the middle square region corresponds to the ground-state wavefunction distributions in the Ge quantum well. In (c), the quantum dot confinement region is marked, corresponding to the localized potential minimum where the HH ground-state wavefunction is confined.

To investigate the effect of lateral device dimensions, we define eight device sizes (Size 1 to Size 8), each a uniformly scaled version of the others, listed in Table 1. This linear scaling preserves geometrical symmetry and allows isolation of dimensional effects without introducing shape anisotropy. The vertical heterostructure profile (along the z direction) remains identical across all device sizes.

Table 1: Lateral dimensions $d_x$ and $d_y$ of simulated devices for various sizes. These values define the extent of the electrostatic confinement region in the in-plane (xy) directions and are scaled proportionally across device sizes to preserve geometric symmetry.

| | | | |
|---|---|---|---|
| Size 1 | 170.0 nm | 162.0 nm | 1.0 |
| Size 2 | 190.4 nm | 181.4 nm | 1.1 |
| Size 3 | 204.0 nm | 194.4 nm | 1.2 |
| Size 4 | 221.0 nm | 210.6 nm | 1.3 |
| Size 5 | 238.0 nm | 226.8 nm | 1.4 |
| Size 6 | 255.0 nm | 243.0 nm | 1.5 |
| Size 7 | 272.0 nm | 259.2 nm | 1.6 |
| Size 8 | 289.0 nm | 275.4 nm | 1.7 |

### II.II Theoretical Calculations

All simulations were performed using the QTCAD software framework.[12] The electronic structure is modeled using a multiband k·p Luttinger–Kohn (L-K) Hamiltonian, incorporating the four top valence subbands: two HH and two LH states. Strain effects due to lattice mismatch in the $Si_{0.2}Ge_{0.8}$/Ge heterostructure are treated using the Bir–Pikus formalism, which modifies the L-K Hamiltonian to account for strain-induced band splitting and mixing. Material parameters are taken either directly from QTCAD's built-in material database or interpolated using its alloy parameter interpolation utility.

The calculation begins with the solution of the Poisson equation to compute the electrostatic potential resulting from gate voltages, band-edge discontinuities, and charge distributions. This potential is then used as input to solve the Schrödinger equation for the quantum-confined states in the Ge quantum well.

Once convergence is reached, the final electrostatic potential is used to solve the Schrödinger equation again in a restricted region corresponding to the quantum dot (as defined by the gate geometry). The restriction to a predefined central dot region is necessary due to the solver's design, which assumes a single global confinement minimum. Solving across the full device would introduce spurious solutions in regions outside the intended dot confinement area.

The resulting eigenfunctions and eigenvalues characterize the confined HH-like qubit states. From these, we extract the spatial profile of the wavefunction, compute the anisotropic g-tensor, and evaluate coherence metrics such as $T_1$ and $T_2^*$ .

A full theoretical overview of the Luttinger–Kohn and Bir–Pikus models, along with a walkthrough of the numerical methodology and governing equations implemented in QTCAD, is also provided in the Appendices A-C.

## III. Results and Discussion

### III.I Wavefunction Confinement Study

In this section, we examine how the lateral dimensions of the device and the applied gate bias influence the confinement properties of the hole ground-state wavefunction. To make the geometric and electrostatic effects more realistic, all simulations are performed assuming relaxed barriers and a compressively strained QW and in an external, uniform magnetic field with a magnitude of 0.1 T. Under these conditions, the computed wavefunctions are dominated by HH character, consistent with our prior results[13], which analyzed the strain dependence of HH–LH mixing in $Si_{0.2}Ge_{0.8}$/Ge/$Si_{0.2}Ge_{0.80}$ heterostructures.

Our analysis focuses on two aspects of spatial confinement: the size of the wavefunction, which quantifies the lateral extent of confinement and is defined here as four times the standard deviation of the wavefunction probability density along each direction, and the average position of the wavefunction, which measures any displacement of the confined state from the geometric center of the quantum dot. Together, these metrics provide insight into both the confinement strength and the symmetry of the electrostatic potential under different device geometries and biasing conditions.

Simulations are performed for all device sizes listed in Table 1 using gate-bias configurations defined by a characteristic voltage $V$ applied to the upper and side confinement gates. In all cases, the bottom-gate bias is fixed at $V + 0.2$ V. Figure 2(a) shows the dependence of the wavefunction position and shape on device size. As the device size increases, the wavefunction shifts upward, away from the bottom gate.

Figure 2(b) shows how the wavefunction evolves under three representative gate-bias configurations. In these cases, 0.4, 0.64, and 0.88 V are applied to the top and side gates, while the bottom-gate voltage remains 0.2 V higher in each case. As the gate voltage increases, the wavefunction shifts away from the top gate and becomes more asymmetric. As discussed in the next section, increasing the gate voltage modifies the confinement potential from a more symmetric to a more asymmetric profile. This enables us to examine how the wavefunction spatial distribution responds to electrostatic tuning of the lateral confinement landscape, as shown in Figure 3. These results illustrate the interplay between gate-defined potential shaping and quantum confinement, which is essential for achieving robust and spatially localized spin qubit states in planar Ge devices.

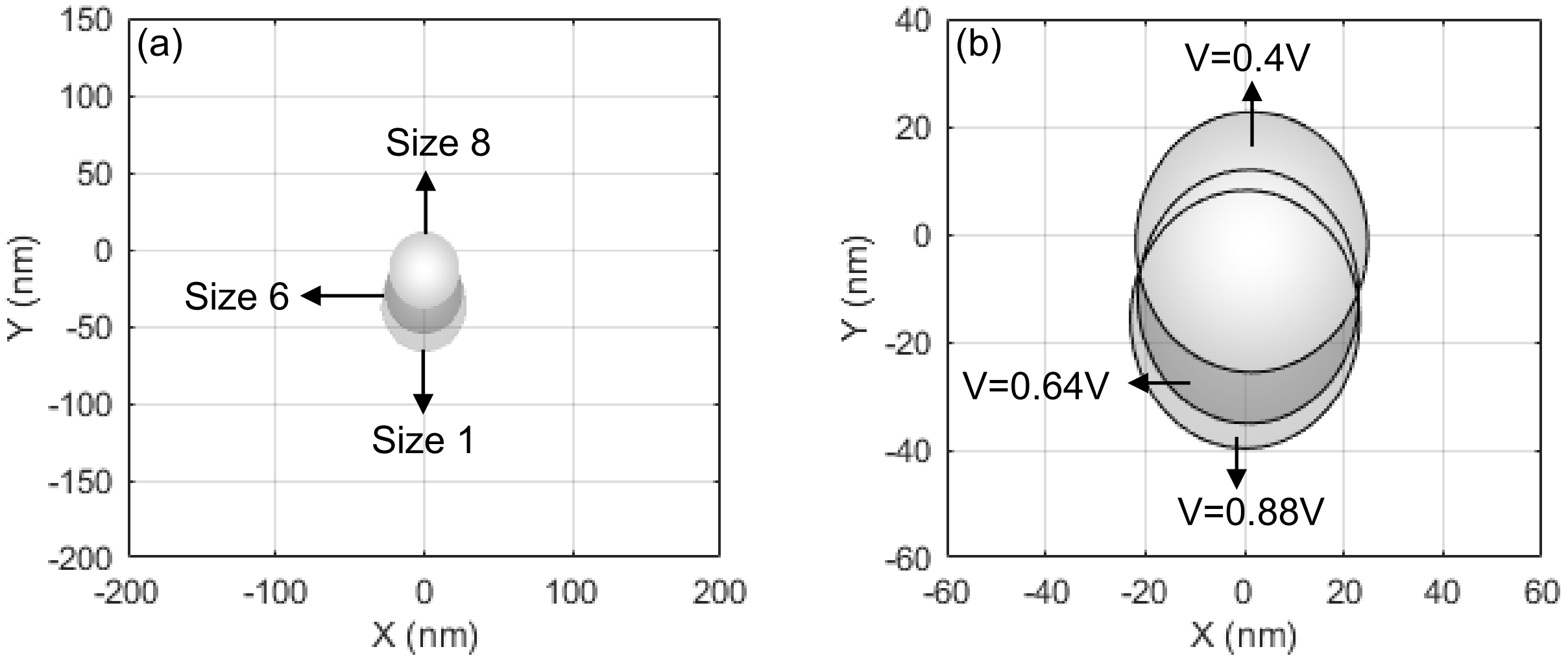

Figure 2: Lateral (xy-plane) spatial distributions of the ground-state hole wavefunction in the Ge quantum well for (a) different device dimensions and (b) different gate-bias configurations. In panel (a), all wavefunctions are calculated for a fixed bias configuration of 0.4 V applied to the side and top gates and 0.6 V applied to the bottom gate. In both panels, the shaded regions denote the first standard deviation of the wavefunction probability density, providing a quantitative measure of the spatial extent and in-plane anisotropy of the confined state.

### III.II Effect of Device Size on Wavefunction Spatial Distribution

Figure 3 illustrates how the valence-band edge (VBE) evolves with device dimensions. In the lateral plane, both the x- and y-direction VBE profiles become steeper as the device size increases. This produces a deeper and more localized in-plane confinement potential. At the same time, the potential minimum becomes increasingly displaced toward the bottom gate, indicating that the larger devices develop a stronger in-plane asymmetry near the center of the semicircular gate pattern formed by the upper three confinement gates. In contrast, the dependence of the out-of-plane VBE on device size is much weaker, appearing mainly as a modest change in the magnitude and slope of the vertical confinement profile. This weaker sensitivity is expected because the applied voltages are used primarily to define lateral confinement rather than to strongly tune the vertical potential in the manner of a dedicated plunger-gate configuration.

The change in the electrostatic potential landscape with device size, both in-plane and out-of-plane, has a strong impact on the wavefunction position and extent, as shown in Figure 5(a) and (b). As the spacing between the confinement gates increases, the direct lateral squeezing of the dot is reduced. However, this is accompanied by a reduction in the effective out-of-plane electric field $E_z$ at the wavefunction location, as shown later in Figure 6(a). The weaker vertical field allows the wavefunction to shift upward in the $+z$ direction with increasing device size [Figure 5(a)]. Because the wavefunction moves farther from the lower interface and into a different part of the electrostatic landscape, the effective lateral confinement experienced by the hole is modified. In our geometry, this results in an overall reduction in the in-plane wavefunction size, reflected in smaller $d_x$ and $d_y$.

The detailed trend of this size reduction is influenced by the asymmetric gate layout. As the device size increases, the in-plane potential tilts progressively away from the central semicircular confinement region, causing the wavefunction to shift relative to the geometric center of the dot. This displacement partially relaxes the lateral confinement and competes with the effect of the vertical shift. As a result, the reduction in wavefunction size does not continue indefinitely, but instead tends toward saturation at larger dimensions. Figure 5(a) further shows that the average x position of the wavefunction remains nearly unchanged with device size, consistent with the approximate symmetry of the confinement potential along x. In contrast, the y position shifts approximately linearly, reflecting the increasing asymmetry of the potential toward the bottom gate.

These changes in wavefunction size and position are particularly important because they directly affect the local asymmetry of the confinement potential and the effective vertical field $E_z$ experienced by the confined hole. Both quantities enter the Rashba spin-orbit interaction and therefore enable device-size-dependent modulation of SOC strength and anisotropy[25]. This provides one of the main links between geometric design and spin dynamics in these structures, as discussed in the following sections.

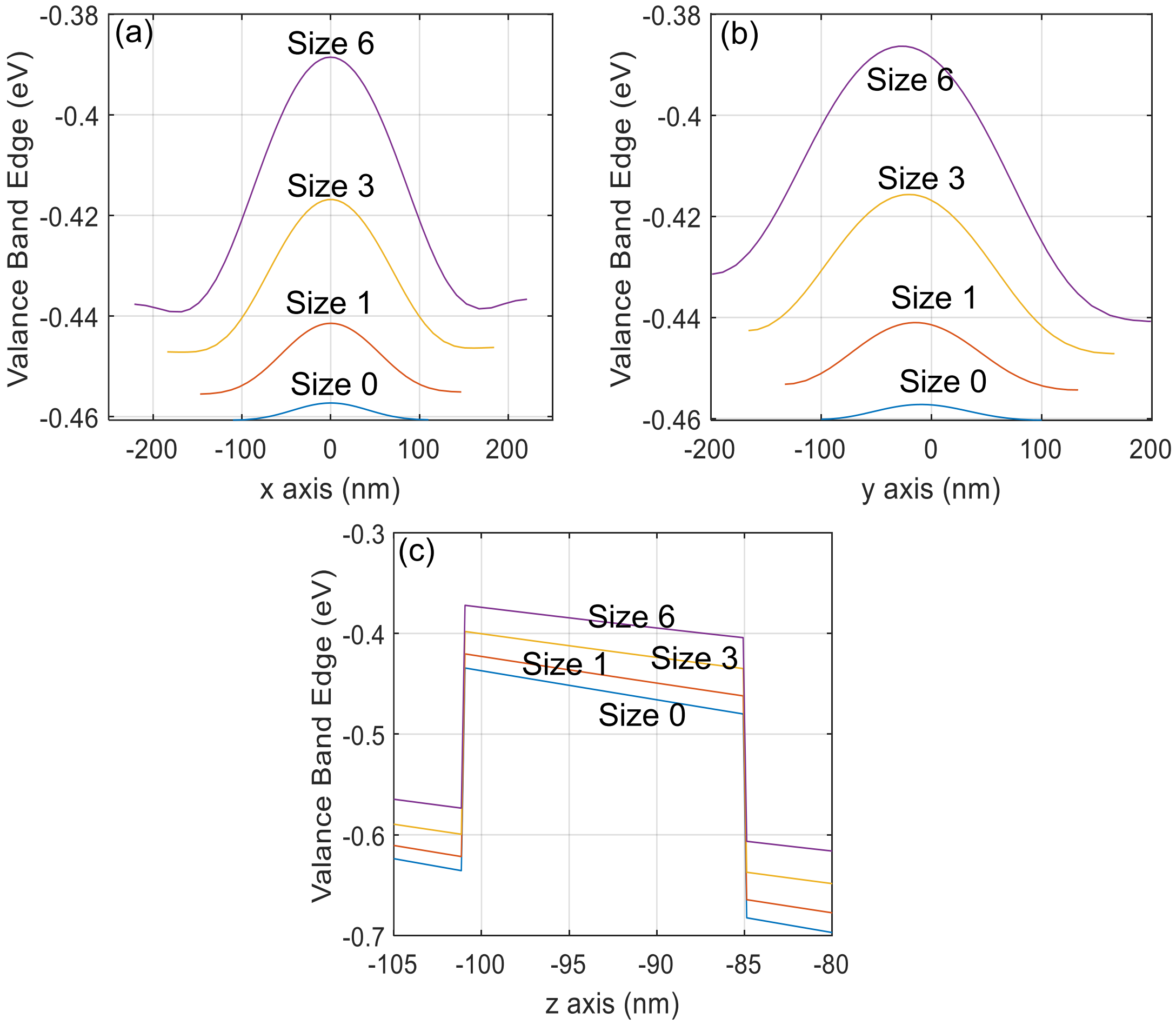

Figure 3: Valence-band edge profiles, in eV, for Size 0, 1, 3, and 6 devices under the 0.7/0.9 V gate-bias configuration, plotted along the (a) x, (b) y, and (c) z directions. The in-plane profiles in (a) and (b) illustrate how lateral confinement and asymmetry evolve with device size, while (c) shows the comparatively weaker size dependence of the vertical confinement along the growth direction.

### III.III Effect of Varying Gate Voltage Bias on Wavefunction Spatial Distribution

Figure 4 illustrates the effect of gate bias on the VBE along the x, y, and z directions. In contrast to the trend observed with device size, increasing the applied gate bias makes the in-plane VBE profiles progressively less steep, indicating weaker lateral confinement. These changes in the electrostatic landscape alter both the symmetry and localization of the wavefunction and therefore are expected to affect the spin–orbit coupling and, consequently, the g-tensor, as discussed in the next section.

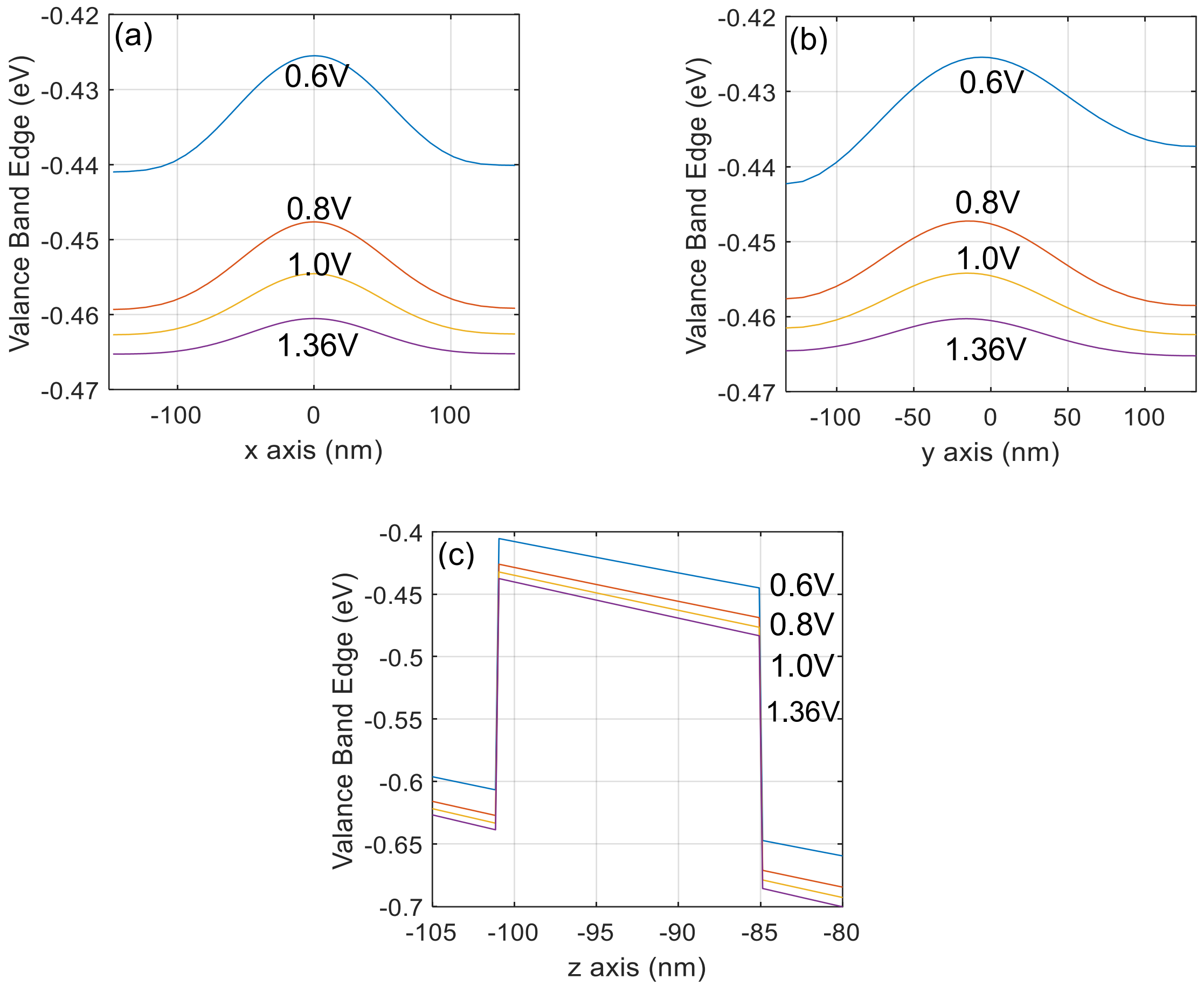

Figure 4: Valence-band edge profiles, in eV, for the Size 1 device under gate-bias configurations of 0.6, 0.8, 1.0, and 1.36 V on the upper and side gates, with the bottom gate biased 0.2 V higher in each case. The band-edge profiles are plotted along the (a) x, (b) y, and (c) z directions to illustrate the bias dependence of the lateral and vertical confinement potential.

The corresponding evolution of the wavefunction position and size is shown in Figure 5(c) and (d). As the gate bias increases, the x position remains nearly unchanged, consistent with the approximate left-right symmetry of the confinement potential. In contrast, the y position [Figure 5(c)] exhibits a pronounced nonlinear dependence on bias. At lower bias, the wavefunction position changes only weakly, but near a threshold around 0.6 V it undergoes a sharp downward shift of nearly 10 nm toward the bottom gate. Beyond this transition, the y position continues to move downward more gradually and approximately linearly with increasing bias. A similar but much smaller trend is observed in the z direction, where the wavefunction shifts downward toward the lower interface by roughly 0.5 nm over the bias range considered. The in-plane wavefunction size also evolves non-monotonically: it initially decreases with bias, corresponding to stronger localization, and then increases significantly once the wavefunction moves into a different confinement regime.

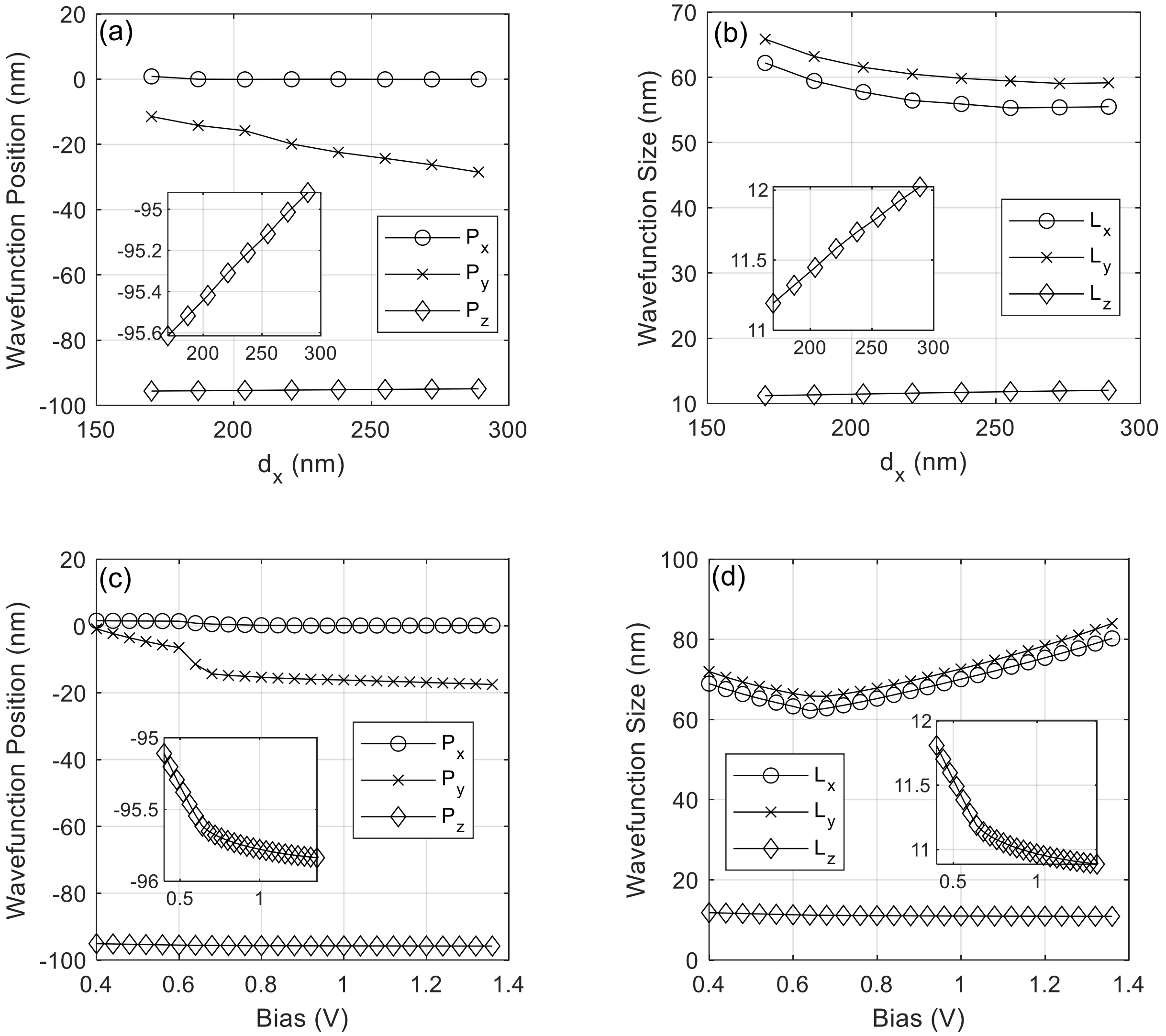

Figure 5: Components of the wavefunction position (P) and spatial extent (L) of the heavy-hole ground state as functions of device dimension and gate bias. The x (circles), y (crosses), and z (diamonds) components are shown separately. (a) Position versus device dimension. (b) Size versus device dimension. (c) Position versus gate bias. (d) Size versus gate bias. Insets highlight the z-component on an expanded scale due to its smaller variation. For (a) and (b), the bias is fixed at 0.64/0.84 V, while for (c) and (d), the device size is fixed to Size 1 (Table 1).

The origin of this behavior lies in the asymmetric gate geometry combined with the biasing scheme. Although the voltage applied to the upper and side gates is increased uniformly, the bottom gate is always biased 0.2 V higher, so the confinement potential becomes increasingly skewed toward the bottom gate as the overall bias increases. This progressively shifts the wavefunction downward and changes the character of the local confinement. To clarify this evolution, it is useful to divide the behavior into three bias regimes.

In Regime I, from about 0.4 to 0.6 V, the wavefunction remains localized within the semicircular confinement region defined mainly by the upper three barrier gates. In this regime, increasing the bias strengthens the local confinement, leading to a roughly monotonic decrease in the lateral wavefunction size. At the same time, the stronger local electrostatic field also pulls the wavefunction slightly downward in both y and z.

At the onset of Regime II, near 0.6 V, the minimum of the confinement potential begins to move out of the local semicircular region. The wavefunction then starts to transition into a broader confinement region formed by the bottom gate, the parallel sections of the side gates, and the residual potential from the upper three gates. Because the wavefunction is no longer confined primarily by the more localized semicircular potential, its y position shifts rapidly toward the bottom gate. This transition occurs over a relatively narrow bias interval, approximately 0.6 to 0.65 V, and is accompanied by a strong change in wavefunction size and anisotropy.

In Regime III, above about 0.65 V, the wavefunction has largely moved into the new confinement configuration. As it approaches the bottom gate, the electrostatic influence of the bottom gate becomes increasingly balanced by the combined effect of the upper three gates. As a result, the shift in y position begins to saturate and changes only gradually with further increase in bias. The z position follows a similar trend: in Regime I the increasing bias enhances the effective vertical field and drives the wavefunction downward more strongly, whereas in Regimes II and III the sensitivity of the z position to bias is reduced because the wavefunction is no longer confined entirely within the original semicircular potential region.

Overall, these results show that gate bias does not merely tune the strength of confinement, but can qualitatively reconfigure the electrostatic confinement landscape experienced by the hole, transferring the wavefunction between the semicircular and approximately rectangular confinement regions. This transition between distinct geometric confinement regimes changes both the local potential asymmetry and the effective vertical field $E_z$ experienced by the wavefunction, two quantities that directly influence Rashba spin–orbit coupling and g-tensor anisotropy.

### III.IV Directional g-Factor Components and Size Dependence

Both device dimensions and gate-bias variations modify the Rashba spin-orbit coupling and, consequently, the g-tensor of the heavy-hole ground state through changes in the confinement potential. To quantify these effects, we introduce an effective out-of-plane electric field, $E_{z,\mathrm{eff}}$, obtained by averaging the local electric field over the full wavefunction distribution, following the approach of Winkler[25]:

$$E_{z,eff} = \int d_V (|\Psi_{HH1}|^2 + |\Psi_{HH2}|^2 + |\Psi_{LH1}|^2 + |\Psi_{LH2}|^2) E_z \qquad (1)$$

Here, HH1, HH2, LH1, and LH2 denote the four components of the J = 3/2 basis, namely, $|\, 3/2, 3/2\rangle$, $|\, 3/2, -3/2\rangle$, $|\, 3/2, +1/2\rangle$, and $|\, 3/2, -1/2\rangle$, respectively, and $V$ is the quantum-dot region over which the wavefunction is defined. Since $E_{z,\mathrm{eff}}$ is determined

entirely by the electrostatic potential and the hole wavefunction, it is independent of the direction of the applied uniform magnetic field. The calculated values of $E_{z,\mathrm{eff}}$ are shown in Figure 6.

Because our 3D simulations include confinement-potential variations in both the out-of-plane ($z$) and in-plane ($xy$) directions under the asymmetric gate-bias configuration, they capture not only the full three-dimensional electrostatic landscape but also the resulting shifts in wavefunction position and size throughout the device. This is important because different in-plane locations experience different sensitivities of $E_z$ to gate bias. As the wavefunction shifts laterally, especially along the $y$direction, it samples regions with different local vertical-field responses. Consequently, the effective field $E_{z,\mathrm{eff}}$ can evolve with gate bias not only because the electrostatic field itself changes, but also because the wavefunction moves to a different part of the device. This effect is reflected in the nonlinear dependence of $E_{z,\mathrm{eff}}$ on gate bias in Figure 6(b), together with the position and size changes shown in Figure 5. In contrast, Figure 6(a) shows that $E_{z,\mathrm{eff}}$ varies approximately linearly with device dimension. Because of this nearly linear size dependence, the bias-dependent curves for different device sizes in Figure 6(b) are approximately offset from one another by a nearly constant shift. These geometry- and bias-induced changes in wavefunction position and extent have direct consequences for the g-factor, as discussed below.

The influence of an external magnetic field on the energy levels of the confined hole spin states is characterized by the effective g-factor through Zeeman splitting. In anisotropic systems such as gate-defined hole quantum dots, the g-factor depends on the direction of the applied magnetic field. For a magnetic field applied along a single coordinate axis, the Zeeman splitting is given by:

$$\Delta E = \mu_B g_i B_i, \qquad i = x, y, z \tag{2}$$

where $\mu_B$ is the Bohr magneton, $B_i$ is the magnetic-field component along the $i$ axis, and $g_i$ is the corresponding principal component of the effective g-tensor.

More generally, assuming a diagonal g-tensor and neglecting off-diagonal spin mixing, the magnitude of the Zeeman splitting for an arbitrary magnetic-field direction can be written as

$$\Delta E = \mu_B |\boldsymbol{g}.\boldsymbol{B}| = \mu_B \sqrt{g_x^2 B_x^2 + g_y^2 B_y^2 + g_z^2 B_z^2} \tag{3}$$

This expression is useful for describing the anisotropic magnetic response of the qubit under general field orientations. In our simulations, we extract the individual components $g_x$, $g_y$, and $g_z$ by applying magnetic fields independently along the x, y, and z axes and calculating the resulting energy splitting of the heavy-hole ground state. These g-tensor components determine the magnetic response of the qubit and are central to both its static properties and its electrically tunable spin dynamics.

We use these relations to evaluate the g-tensor of the heavy-hole ground state in our devices. In particular, we focus on the in-plane components $g_x$ and $g_y$, corresponding to magnetic fields applied along the x and y directions, and the out-of-plane component $g_z$, corresponding to a field along the z direction. These quantities define the anisotropic spin response of the qubit and directly affect its Larmor frequency, sensitivity to magnetic and electric noise, and coherence properties. Understanding how the g-tensor evolves with device geometry and gate bias is therefore essential for designing electrically tunable and robust spin qubits. In addition, gate-dependent modulation of the g-tensor provides a route toward electrically driven spin control through g-tensor modulation resonance.

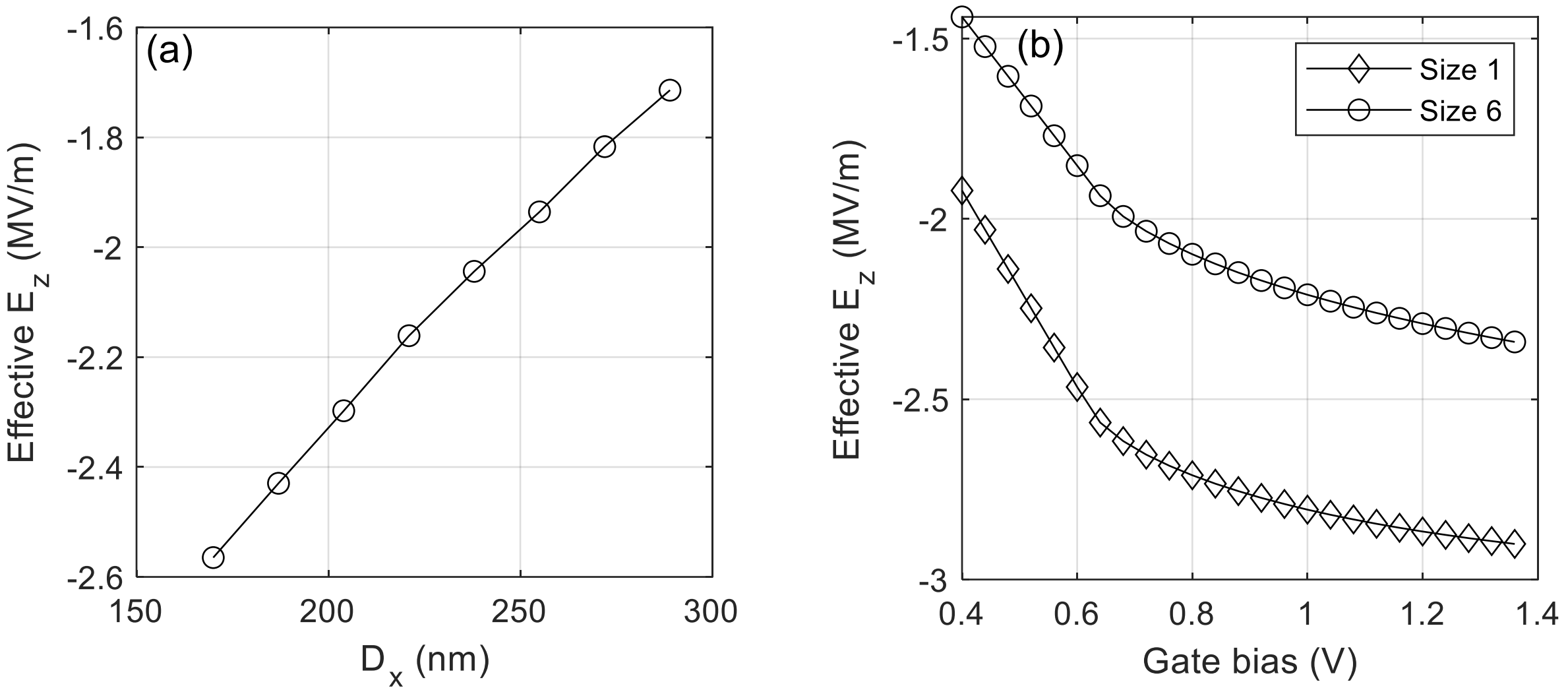

Figure 6: (a) Effective out-of-plane electric field $E_{z,\mathrm{eff}}$ as a function of device dimension; the corresponding device sizes are listed in Table 1. (b) Effective out-of-plane electric field $E_{z,\mathrm{eff}}$ as a function of gate bias for the Size 1 (circles) and Size 6 (diamonds) devices. Here, $E_{z,\mathrm{eff}}$ is calculated by integrating the z component of the electric field over the full device volume, weighted by the ground-state wavefunction probability density.

To characterize the dependence of the in-plane and out-of-plane g-factors on device dimensions, we extracted $g_x$, $g_y$, and $g_z$ across the eight device sizes listed in Table 1 while holding the applied gate bias fixed at the 0.4/0.6 V and 0.64/0.84 V configurations. As shown in Figure 7, the in-plane g-factor components $g_x$and $g_y$exhibit an overall decrease in magnitude as the lateral device dimensions increase. This trend is consistent with the evolution of the wavefunction confinement shown in Figure 5(b): as the dot becomes larger, the spatial distribution of the confined state changes in a way that reduces the effective HH–LH mixing relevant for in-plane Zeeman response, thereby lowering the spin susceptibility to magnetic fields applied in the plane. In contrast, the out-of-plane g-factor $g_z$ shows the opposite behavior and increases with device size. This enhancement is consistent with the increasingly heavy-hole-like character of the ground state in larger devices, which strengthens the magnetic response along the growth ($z$) direction. These trends are consistent with previous analyses of the effect of confinement on g-tensor anisotropy.[14,15] It should be noted, however, that in the present structures the in-plane wavefunction dimensions $L_x$ and $L_y$ evolve in a nearly similar manner under both size and bias variation, so the degree of in-plane ellipticity remains approximately unchanged. In this sense, the dots retain a response qualitatively similar to that of a nearly circular confinement potential.

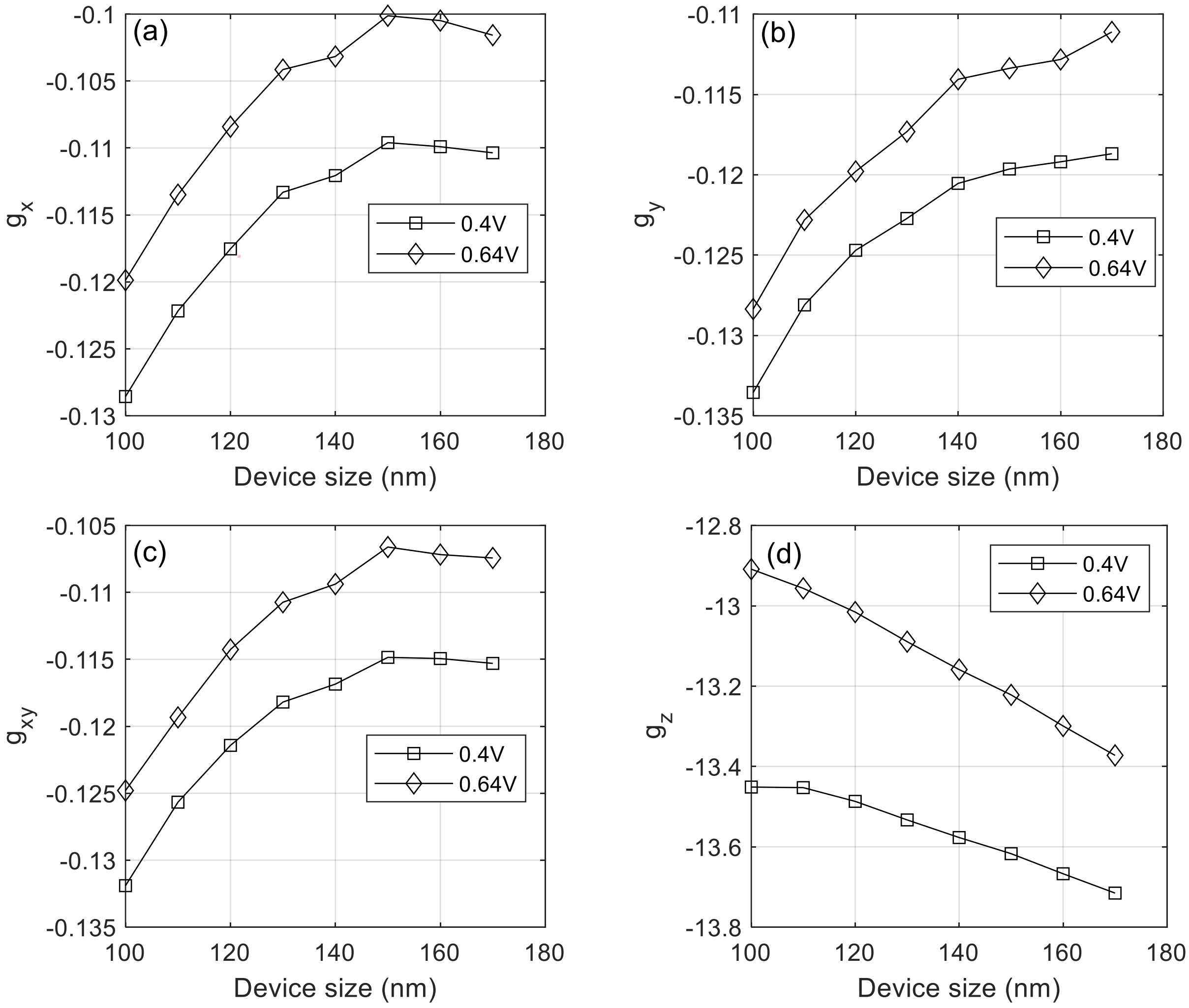

Figure 7: Dependence of the effective g-factor components on device lateral dimensions: (a) $g_x$, (b) $g_y$, (c) $g_{xy}$ (in-plane diagonal direction) and (d) $g_z$, corresponding to magnetic fields applied along the x, y, x=y (45° in-plane), and z (growth) directions, respectively.

The extracted values of $g_x$ and $g_y$ in the dimensional analysis are more susceptible to numerical noise than in the bias analysis. This is partly due to the smaller Zeeman splitting associated with in-plane magnetic fields, and partly due to mesh-dependent numerical differences between simulations carried out for different device sizes. By contrast, the bias-dependent data are obtained using the same 3D mesh for a given device, which reduces numerical scatter. For this reason, we focus primarily on the overall dimensional trends of $g_x$ and $g_y$, rather than on their exact absolute values, and we do not use the dimension-dependent in-plane g-factors from Figure 7 in calculations that are highly sensitive to numerical derivatives.

We next analyze the sensitivity of the g-tensor to gate bias by correlating the bias-induced changes in the electrostatic potential with the extracted g-factor components, as shown in Figure 8. Under asymmetric gate-bias modulation, the combined changes in wavefunction position and size give rise to strongly nonlinear bias regions in which the slope of the g-factor changes abruptly. These regions correspond directly to the rapid positional transition discussed in Section III.III.

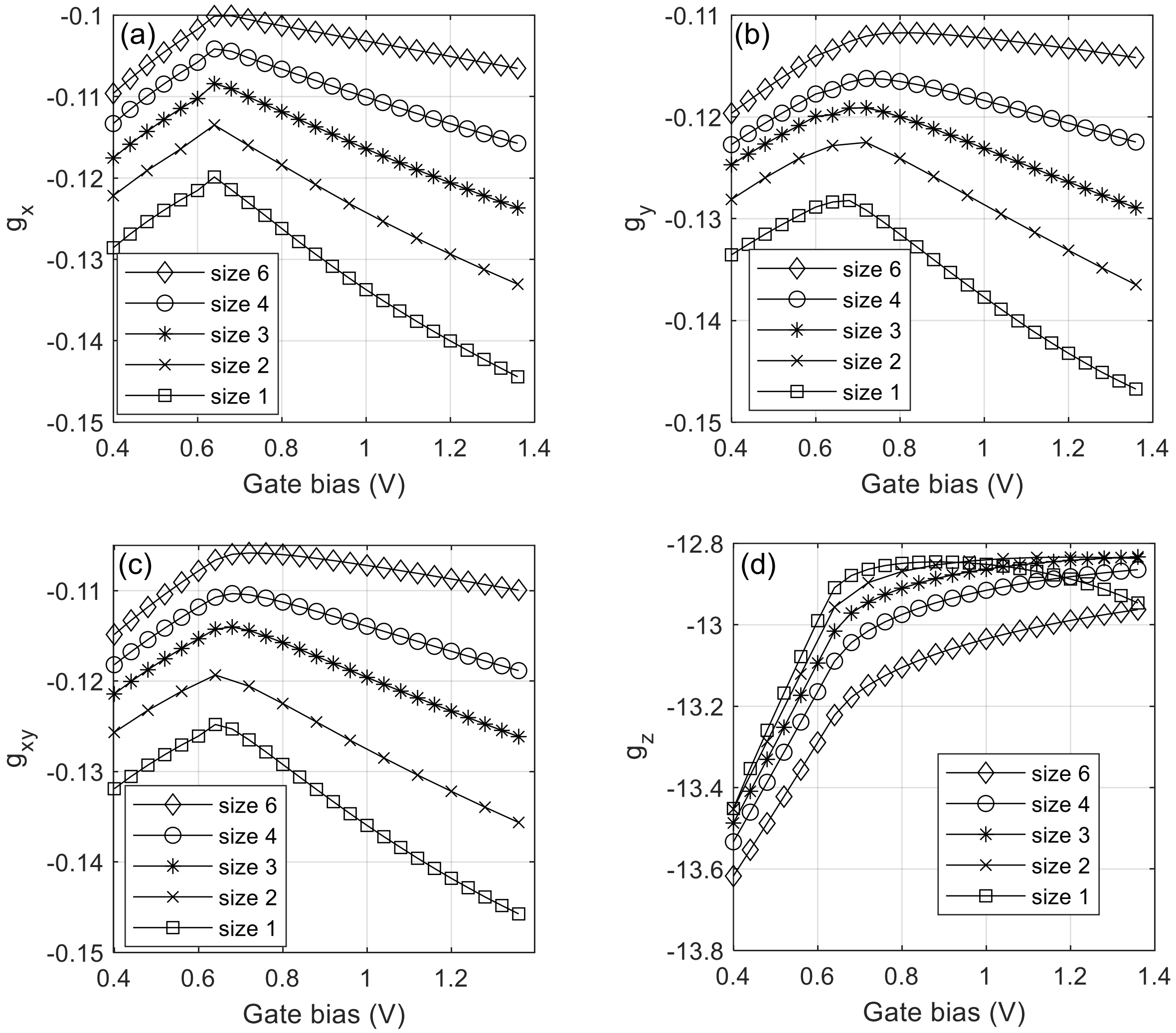

Figure 8: Dependence of the effective g-factors on confinement-gate bias for the Size 1, 2, 3, 4, and 6 devices. (a) $g_x$, (b) $g_y$, (c) $g_{xy}$ and (d) $g_z$, corresponding to magnetic fields applied along the x, y, x=y, and z (growth) directions, respectively. The bias dependence reflects the sensitivity of the g-tensor to electrostatic confinement through its influence on the effective vertical electric field, wavefunction position, and HH–LH mixing in the confined hole state. The nonmonotonic behavior and slope changes observed in certain bias ranges are associated with transitions between distinct confinement regimes induced by the asymmetric gate design. This electrostatic tunability of the g-tensor is central to understanding and engineering spin coherence, geometry-induced dephasing sweet spots, and electrically controlled spin manipulation in gate-defined quantum dot qubits.

In Bias Regime I, identified previously from the position- and size-versus-bias analysis, the wavefunction remains confined primarily within the semicircular potential region defined by the upper three gates. In this regime, increasing the barrier-gate bias strengthens the local confinement and reduces the magnitude of the g-factor. In Bias

Regime III, the wavefunction is localized mainly in the lower, approximately rectangular confinement region. There, increasing the gate bias shifts the wavefunction laterally away from the semicircular region and vertically downward, which modifies the confinement environment and leads to an increase in the g-factor magnitude. In Bias Regime II, the wavefunction lies between these two confinement configurations and undergoes a transition between the two opposite trends observed in Regimes I and III. As a result, the derivative of the g-factor with respect to bias must pass through zero in this regime. This forms the origin of the dephasing sweet spot associated with this asymmetric gate design.

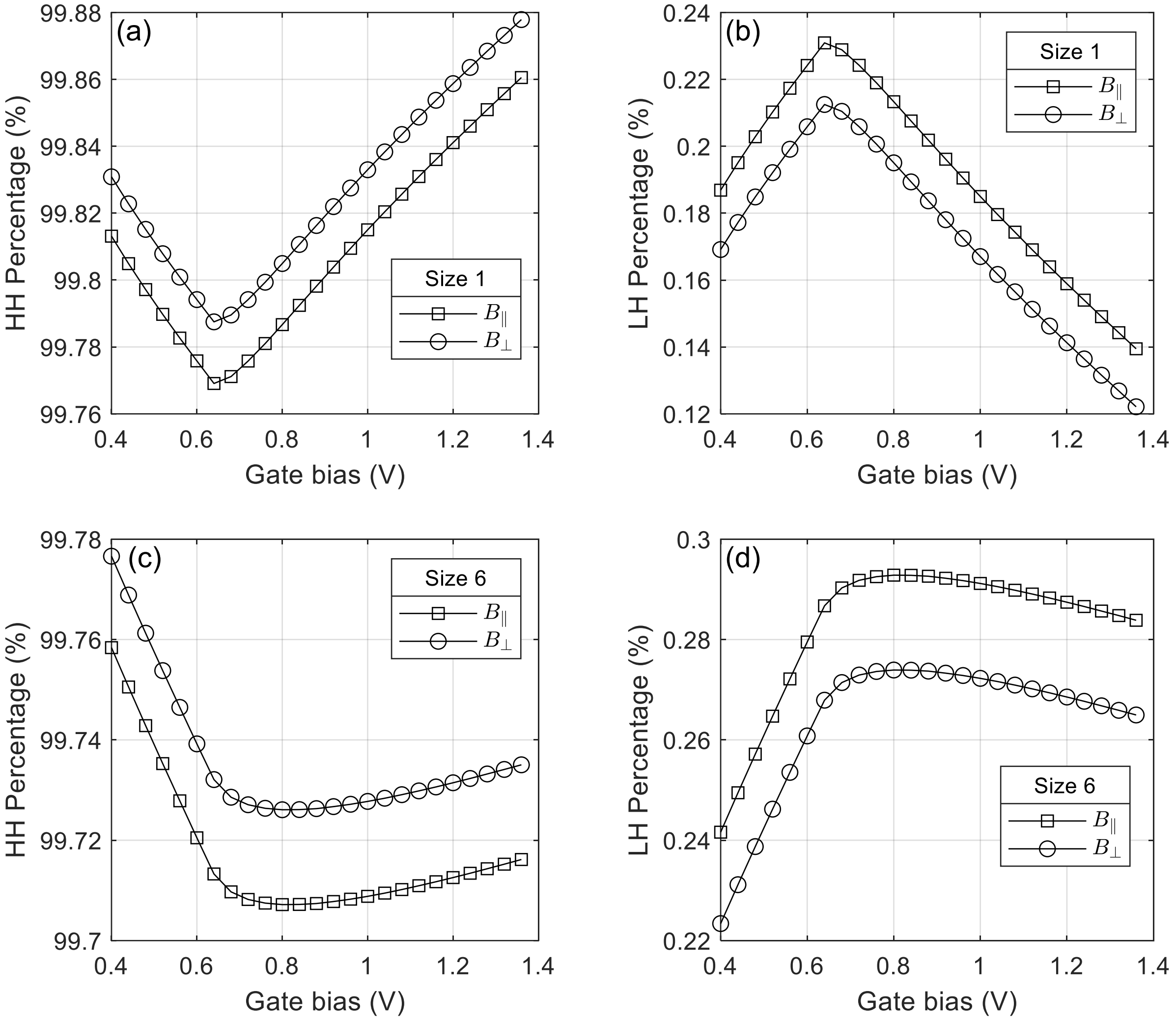

Figure 9: HH and LH band weights, expressed as percentages, of the ground-state wavefunction as functions of gate bias. Panels (a) and (b) show the HH and LH weights for the Size 1 device, respectively, while panels (c) and (d) show the HH and LH weights for the Size 6 device, respectively. Squares correspond to an in-plane magnetic field, and circles correspond to an out-of-plane magnetic field. The bias dependence of the HH/LH composition reflects changes in confinement and wavefunction redistribution that accompany the observed evolution of the effective g-factor.

An additional observation is that the g-factor transition between Bias Regimes I and III becomes less abrupt as the device dimension increases. As the gates move farther apart, the potential gradient between the semicircular and rectangular confinement regions is reduced, so the wavefunction shift becomes more gradual and the corresponding g-factor change becomes smoother. This supports the conclusion that the sweet spot originates from wavefunction motion within an intentionally engineered in-plane gate landscape, and it suggests that the width of the sweet-spot region may be tunable through gate spacing and related confinement engineering. To further illustrate this point, we plot the heavy-hole and light-hole band weights of the ground-state wavefunction as functions of gate bias. As shown in Figure 9, the change in g-factor behavior is accompanied by a corresponding change in the HH/LH composition of the wavefunction. This is consistent with the structure of the Zeeman Hamiltonian used in the model, whose effective response depends strongly on the relative HH and LH contributions to the envelope-function state.

### III.V Spin Orbit Coupling

#### III.V.I Rashba and Dresselhaus Spin-Orbit Interactions

The spin degeneracy of electronic states in semiconductors originates from the combined presence of time-reversal and spatial inversion symmetry in the crystal Hamiltonian. In the absence of external perturbations, these symmetries prohibit spin splitting at zero magnetic field. To lift this degeneracy, at least one of these symmetries must be broken. In gate-defined quantum dots, this is typically achieved through the breaking of spatial inversion symmetry, which gives rise to SOC and enables electrically tunable spin splitting even at zero magnetic field.

For quasi-two-dimensional SiGe/Ge quantum wells, such as those considered in this work, SOC is dominated by structural inversion asymmetry (SIA), commonly referred to as the Rashba effect. This arises because both silicon and germanium possess bulk inversion symmetry due to their diamond crystal structure, which suppresses bulk inversion asymmetry (BIA), the mechanism responsible for Dresselhaus SOC. As a result, Dresselhaus-type SOC, prominent in III–V semiconductors such as GaAs and InAs, is negligible in Ge-based systems, allowing Rashba SOC to be treated as the leading mechanism governing spin splitting.

Rashba SOC originates from asymmetries in the confining potential, which may arise intrinsically from band offsets in the heterostructure or be externally imposed through gate-induced electric fields. A key feature of this interaction is its electrical tunability: the strength of the Rashba coupling, and therefore the spin splitting, g-factor anisotropy, and effective spin–orbit fields, can be controlled via gate voltages. This property underpins all-electrical spin manipulation techniques such as electric dipole spin resonance (EDSR) and g-tensor modulation.

From a theoretical perspective, Rashba SOC in semiconductor nanostructures is typically treated within the envelope function approximation (EFA) using multiband $\mathbf{k} \cdot \mathbf{p}$ Hamiltonians. Within this framework, SIA-induced SOC can either be introduced explicitly through symmetry-derived Rashba terms or incorporated implicitly via quasi-degenerate perturbation methods such as Löwdin partitioning.[17] In full-band models, Rashba effects are embedded in the underlying band parameters. In reduced models, such as the 4×4 or 8×8 Luttinger–Kohn Hamiltonians, the influence of remote bands must be captured through effective coupling terms derived from higher-order corrections.

In particular, even within a 4×4 Luttinger–Kohn Hamiltonian that explicitly includes only heavy-hole and light-hole states, the effects of SIA and remote-band coupling can be incorporated through Löwdin partitioning. This allows the model to retain key physical features, including electric-field-dependent g-tensor modulation and spin–orbit-driven transitions, while maintaining computational efficiency suitable for large-scale device simulations.

### III.V.II Effective Modeling of Rashba SOC Using the Four-Band Luttinger–Kohn Hamiltonian

In this study, the spin–orbit coupling behavior in the gate-defined Ge quantum dots is modeled using QTCAD's implementation of a four-band L-K Hamiltonian, which captures the topmost four valence subbands (two heavy-hole and two light-hole states). This model includes appropriate boundary conditions to reflect the band offsets and material discontinuities in the $Si_{0.2}Ge_{0.8}$/Ge heterostructure. While the four-band L-K Hamiltonian is widely used to model hole systems in semiconductors, it is not strictly equivalent to the effective 4×4 Hamiltonian derived via Löwdin partitioning from the full infinite-band envelope function approximation (EFA). As a result, it does not include all remote-band contributions that affect spin–orbit coupling, and the Rashba interaction is not explicitly parameterized.

Nevertheless, our simulations show that the four-band L-K model effectively reproduces the key Rashba physics relevant to gate-defined Ge quantum dots. To evaluate this, we extract an effective Rashba coefficient $v_3$ from our gate-dependent g-factor simulations and compare it to theoretical values obtained from analytical models that explicitly include Rashba-type terms.

First, we use Equation (4) from Hetényi, Bosco, and Loss[16], which relates the Rashba energy scale D to $v_3$ through the expression:

$$D = m^* v_3^2 \frac{\hbar^4}{L_z^4} \mu \tag{4}$$

Using representative parameters, D = $5\mu eV$, $m^* = 0.346 m_0$, $L_z = 15.8\, nm$, $\mu = 0.6$, we obtain:

$$v_3 = 1.2 \times 10^{18}\ m^3/(eV^2.s^3) \tag{5}$$

Second, we independently estimate $v_3$ using a second-order perturbative analysis of the Rashba Hamiltonian:

$$H_{SO} = v_3 p^3 \sigma_y \tag{6}$$

From standard perturbation theory, the Rashba-induced energy splitting is given by:

$$\delta E \sim \sum\nolimits_n \frac{|\langle n|H_{so}|0\rangle|^2}{E_0 - E_n} \sim \frac{(p^3 v_3)^2}{\Delta_{orb}} \tag{7}$$

Assuming the energy splitting contributes additively to the Zeeman effect, we model the total energy shift as:

$$\Delta E = g_0 \mu_B B + \delta E \approx (g_0 + \delta g^*) \mu_B B \tag{8}$$

with $\delta E \sim \delta g^* \mu_B B$, and the orbital energy scale approximated by:

$$\Delta_{orb} \sim \frac{\hbar^2}{2m^* L_z^2} \tag{9}$$

as a simple analytical estimate. In the actual numerical calculations used to generate the plots in this work, however, $\Delta_{\rm orb}$ is obtained more accurately from a Luttinger–Kohn analysis of the heterostructure along the growth direction, as described in Appendix D. This numerically extracted splitting is then used to evaluate the effective Rashba coefficient entering the relaxation model.

Solving for $v_3$, we obtain the following expression:

$$v_3 \sim \frac{L_z^2}{\hbar^2 \sqrt{2m^*}} \sqrt{\delta g^* \mu_B B} \tag{10}$$

We determine $\delta g^*$ by fitting the simulated g-factor as a function of vertical electric field $E_z$ using a second-order polynomial:

$$g^*(E_z) \approx \; g_0 + \alpha E_z + \; \beta E_z^2 \tag{11}$$

To obtain an approximate analytical estimate of $v_3$, we use the fitted curvature $\delta g^*$ together with the simplified expression for $\Delta_{\rm orb}$ given above. We emphasize that, for the numerical calculations used to generate the plots in this work, $\Delta_{\rm orb}$ is evaluated more accurately from an independent Luttinger–Kohn calculation (Appendix D). Within the present approximation, this procedure gives:

$$v_3 \approx 2.5 \times 10^{18} \; m^3/(eV^2 . s^3) \tag{12}$$

Despite the simplifying assumptions, including approximate treatment of the orbital scale and neglect of higher-order corrections, this value is in reasonable agreement with the analytical estimate. This consistency reinforces the conclusion that the four-band Luttinger–Kohn model, although it does not capture all remote-band contributions, successfully incorporates the leading-order Rashba spin–orbit interaction through its effective band parameters and electric field sensitivity.

In sum, these results validate the use of the QTCAD four-band L-K Hamiltonian for modeling Rashba-type spin–orbit coupling in Ge quantum dot systems. They also demonstrate that gate-modulated g-factors can serve as a viable probe for extracting effective spin–orbit parameters in systems where SOC is strong and electrostatically tunable.

### III.VI Analysis of Dephasing Sweet Spots from Electric-Field-Induced g-Factor Modulation

Understanding and controlling qubit dephasing is central to the performance of spin-based quantum information platforms. In this section, we investigate the dephasing behavior of hole spin qubits in gate-defined $Si_{0.2}Ge_{0.8}$/Ge quantum dots by analyzing how electrostatically tunable g-tensor components couple to low-frequency charge noise. Specifically, we focus on the emergence of *dephasing sweet spots*, bias conditions under which the qubit becomes first-order insensitive to fluctuations in the out-of-plane electric field $E_z$.

To model decoherence, we adopt a well-established framework for 1/f charge noise, assuming a power spectral density of the form $S(f) = A/|f|$, where A = 450 $V^2/m^2$, consistent with measurements in semiconductor heterostructures[17]. The resulting dephasing time $T_2^*$, limited by quasistatic field fluctuations, is calculated using:

$$\frac{1}{T_2^*} = \left|\partial_{E_z}\omega\right|\sqrt{Alog(1/_{2\pi f_{ir}t})} \quad (13)$$

with $f_{ir}$=1 Hz, t=10 µs, and ω defined via:

$$\hbar\omega = \sqrt{\mu_B^2 B_j B_k g_{ij} g_{ki}} \quad (14)$$

This formalism captures dephasing induced by fluctuations of the qubit frequency arising from electrostatic noise in the confinement potential. In our treatment, the key derivative $\partial\omega/\partial E_z$ quantifies the sensitivity of the qubit to fluctuations in the effective out-of-plane electric field experienced by the confined hole, and is evaluated numerically from the $g$-factor-versus-$E_z$ curves extracted in Section III.IV. In this sense, extrema of the $g$-tensor response, or "$E_z$-sweet spots," defined by $\partial\omega/\partial E_z = 0$, correspond to operating points with reduced sensitivity to the dominant vertical component of electrostatic noise. We emphasize, however, that these are not necessarily global sweet spots with respect to all charge-noise mechanisms, since microscopic fluctuators can also generate nonuniform electric-field perturbations and higher-order contributions beyond the present first-order treatment.

As shown in Figure 8, the in-plane g-tensor components ($g_x$, $g_y$, and $g_{xy}$) exhibit strongly nonlinear dependences on gate bias, and therefore on the effective electric field $E_z$. This behavior originates from gate-controlled HH–LH mixing together with the bias-induced redistribution of the wavefunction within the asymmetric confinement potential. The resulting nonmonotonic dependence leads to local extrema in $\omega(E_z)$, and hence to operating points where $\partial\omega/\partial E_z = 0$. Within the first-order approximation of Eq. (13), these points correspond to divergences of $T_2^*$, and therefore define *geometry-induced dephasing sweet spots*. These operating points are shown in Figure 10 for both Size 1 and Size 6 devices. Although the detailed dephasing behavior depends on device geometry, at least one such optimum appears in each device for selected magnetic-field orientations.

It is important to distinguish these geometry-induced dephasing sweet spots from the intrinsic sweet spots predicted by Malkoc, Stano, and Loss[17] for silicon hole-spin qubits. In that work, the sweet spots arise from higher-order corrections to the Luttinger Hamiltonian, which modify the intrinsic electric-field susceptibility of the qubit frequency and can strongly enhance $T_2^*$. Because those higher-order terms are not included in the present four-band Luttinger–Kohn model, the physical origin of the extrema in our calculations is different. Our results instead show that, by combining an asymmetric gate design with a three-dimensional gate-defined confinement landscape, one can intentionally create extrema in the g-tensor through controlled wavefunction motion between distinct electrostatic confinement regions. In this sense, the dephasing optimum found here is not an intrinsic material sweet spot, but an electrostatically engineered one.

This distinction is also relevant when comparing with recent germanium hole-qubit experiments. Hendrickx et al.[8] reported sweet-spot operation associated with the electric tunability of a highly anisotropic heavy-hole g-tensor and the resulting suppression of charge-noise sensitivity. While both their work and ours involve extrema in the electric-field response of the qubit frequency, our model indicates that, in the present device, the dominant mechanism is the intentional reshaping of the confinement potential and the accompanying change in wavefunction position and HH–LH composition, rather than the intrinsic higher-order spin–orbit terms emphasized in the silicon-hole theory.

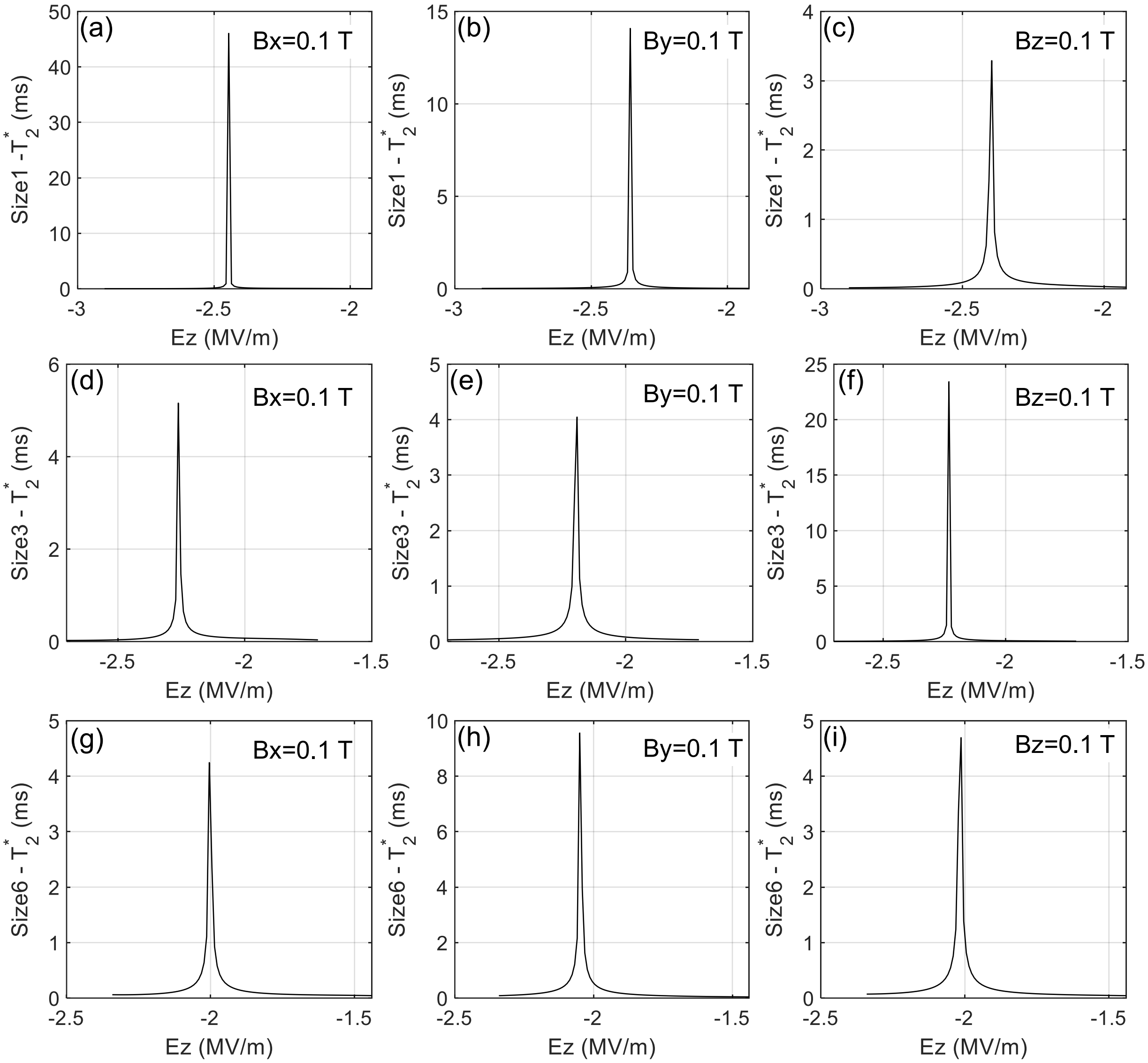

Figure 10: Calculated dephasing time $T_2^*$ as a function of the effective out-of-plane electric field $E_z$(up to an additive constant). The top row shows results for the Size 1 device with magnetic field applied along the x (a), y (b), and $x = y$ (c) directions. The middle row shows the corresponding results for the Size 3 device: x (d), y (e), and $x = y$ (f). The bottom row shows the results for the Size 6 device: x (g), y (h), and $x = y$ (i). Local maxima in $T_2^*$, marked by sharp peaks, correspond to geometry-induced dephasing sweet spots, that is, bias points where the qubit becomes first-order insensitive to vertical electric-field fluctuations because $\partial\omega/\partial E_z = 0$. In the present devices, these optima arise from the nonmonotonic electric-field

dependence of the in-plane g-tensor components, caused by gate-controlled HH–LH mixing together with wavefunction redistribution in the asymmetric confinement potential. Their existence and location are therefore strongly geometry dependent. Although $T_2^*$ formally diverges at these points within the first-order model, they identify realistic operating regions of reduced charge-noise sensitivity.

Our findings are also broadly consistent with the more general sweet-spot concepts explored in silicon spin qubits, although the microscopic mechanisms differ.[17,18,19] In silicon electron qubits, optimal operating points often arise from valley physics, spin-valley hybridization, or engineered magnetic-field gradients rather than from HH–LH mixing. In contrast, the dephasing optima in the present Ge hole system are governed primarily by bias-controlled HH–LH admixture and wavefunction relocation in an asymmetric three-dimensional confinement potential. This makes them especially sensitive to gate geometry and suggests that they can, in principle, be intentionally broadened or shifted through device design.

An important numerical consideration is that, near a geometry-induced dephasing sweet spot, the derivative $\partial\omega/\partial E_z$ approaches zero, and Eq. $(13)$ yields divergent $T_2^*$ values. In practice, this divergence is cut off by higher-order electric-field terms, noise components in other directions, and additional decoherence channels such as spin relaxation. Nevertheless, the location and width of these engineered optima remain physically meaningful, because they identify bias regions where the qubit is substantially less sensitive to charge noise.

These results show that careful design of the gate geometry and bias configuration can create engineered operating points that enhance spin coherence through electrical control of the g-tensor. This provides an additional degree of freedom beyond intrinsic material optimization and makes gate-defined Ge hole spin qubits particularly attractive for scalable, electrically driven quantum architectures.

### III.VIII Gate Engineering of g-Factor Sweet Spots

To further demonstrate the design flexibility of the geometry-induced dephasing sweet spots identified in this work, we investigate the effect of introducing an additional circular accumulation gate at the center of the confinement region. Two accumulation-gate bias values were considered, $-0.1$ V and $-0.3$ V.

The results show that adding the accumulation gate weakens the sharp transition previously induced by the combination of asymmetric biasing and gate geometry. As a consequence, the g-factor sweet spot becomes broader and shifts to higher confinement-gate bias compared with the case without an accumulation gate. This behavior can be understood from the enhanced circular confinement produced by the additional gate: the wavefunction remains more strongly localized in the upper semicircular region, so a larger asymmetry in the barrier-gate bias is required before the wavefunction undergoes its downward shift in the $(-y)$ direction. In other words, the accumulation gate delays the transition between confinement regimes and smooths the associated change in the g-factor. These results suggest that an accumulation gate may provide a practical route for broadening and repositioning the engineered dephasing sweet spot through additional electrostatic control.

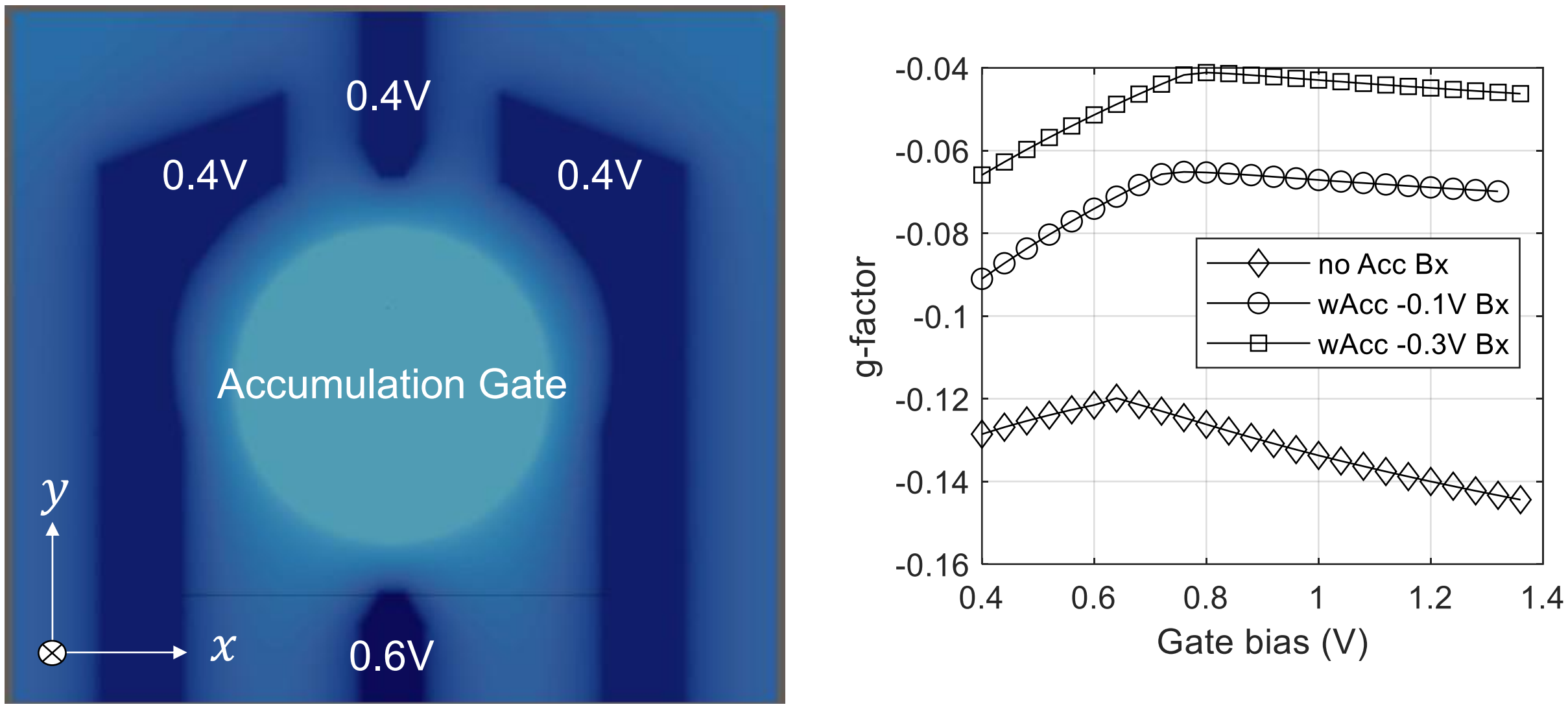

Figure 11: Dependence of the in-plane g-factor on confinement-gate bias in the presence of a circular accumulation gate placed at the center of the semicircular confinement region. Two accumulation-gate bias values are considered: $-0.1$ V and $-0.3$ V. Relative to the case without an accumulation gate, the transition in the g-factor becomes smoother and shifts to higher bias, indicating that the accumulation gate broadens and delays the engineered sweet-spot region by strengthening the central circular confinement.

This result highlights that, unlike intrinsic material sweet spots, the dephasing optima in this system can be deliberately engineered and tuned through gate design, providing a practical pathway toward robust and scalable qubit operation.

### III.IX Phonon-Induced Spin Relaxation and its Dependence on Quantum Dot Geometry

In gate-defined $Si_{0.2}Ge_{0.8}$/Ge quantum dot hole spin qubits under compressive strain, the valence-band ground state is predominantly HH-like, as established in prior theoretical and experimental studies.[13] When hyperfine interactions are weak, either because of isotopic purification or the $p$-orbital character of valence-band holes, the dominant spin-relaxation pathway is phonon-mediated and is enabled by spin-orbit coupling (SOC).

In group-IV semiconductors such as Ge, bulk inversion asymmetry is absent because of the diamond crystal structure, so Dresselhaus SOC is negligible.[20] Rashba-type SOC, arising from structural inversion asymmetry and tunable through gate-induced electric fields, therefore becomes the dominant spin-orbit mechanism in the present system. This provides the basis for SOC-mediated phonon-induced spin relaxation.

We adopt the theoretical model developed by Bulaev and co-workers[2, 21] which describes phonon-induced spin relaxation in HH systems with Rashba SOC. The spin-relaxation rate is given by:

$$W_{1n}^{R} = \frac{\alpha^2 \omega_z^7}{2^8 \hbar^3 \pi^2 \rho \Omega^6} \coth\left(\frac{\hbar\omega_z}{2k_B T}\right) \left(\frac{\omega_+^3}{3\omega_+^3 + \omega_z} - \frac{\omega_-^3}{3\omega_-^3 - \omega_z}\right)^2 \sum_\alpha s_\alpha^{-9} e^{-\frac{\omega_z^2 l^2}{2s_\alpha^2}} I^{(7)} \tag{15}$$

where $\alpha$ is the Rashba spin-orbit coupling coefficient, $\omega_z = g_z \mu_B B/\hbar$ is the Zeeman frequency, $\rho$ is the crystal mass density, $\Omega$ is the effective in-plane confinement frequency, $\omega_\pm$ are the hybrid confinement frequencies defined in the Rashba-mediated relaxation model of Ref. [2], $s_\alpha$ is the sound velocity for phonon branch $\alpha$, and $l$ is the lateral confinement length. The factor $I^{(7)}$ is an angular-radial integral given by:

$$I^{(k)} = \int_0^{2\pi} d\varphi \int_0^{\frac{\pi}{2}} d\vartheta \, sin^k\vartheta \, F^2\left(\frac{\omega_z \, cos\vartheta}{s_\alpha}\right) e^{\frac{\omega_z^2 l^2 cos^2\vartheta}{2 s_\alpha^2}} \left\{ \left(eA_{q\alpha}\right)^2 + \frac{\omega_z^2}{s_\alpha^2}\left[\left(a + \frac{b}{2}\right) \boldsymbol{q_v} \cdot \boldsymbol{d^{q\alpha}} - \frac{3}{2} b q_v^z d_z^{q\alpha}\right]\right\} \quad (16)$$

where $F$ is the form factor associated with the wavefunction extent along the growth direction, extracted from the simulated 3D wavefunction; $A_{q\alpha}$ is the phonon-coupling amplitude; $a$ and $b$ are deformation-potential constants; $\mathbf{q}_v$ is the normalized phonon wave vector; $\mathbf{d}^{q\alpha}$ is the phonon polarization vector; and $d_z^{q\alpha}$ is its z component. All material parameters are taken from Ref. [13].

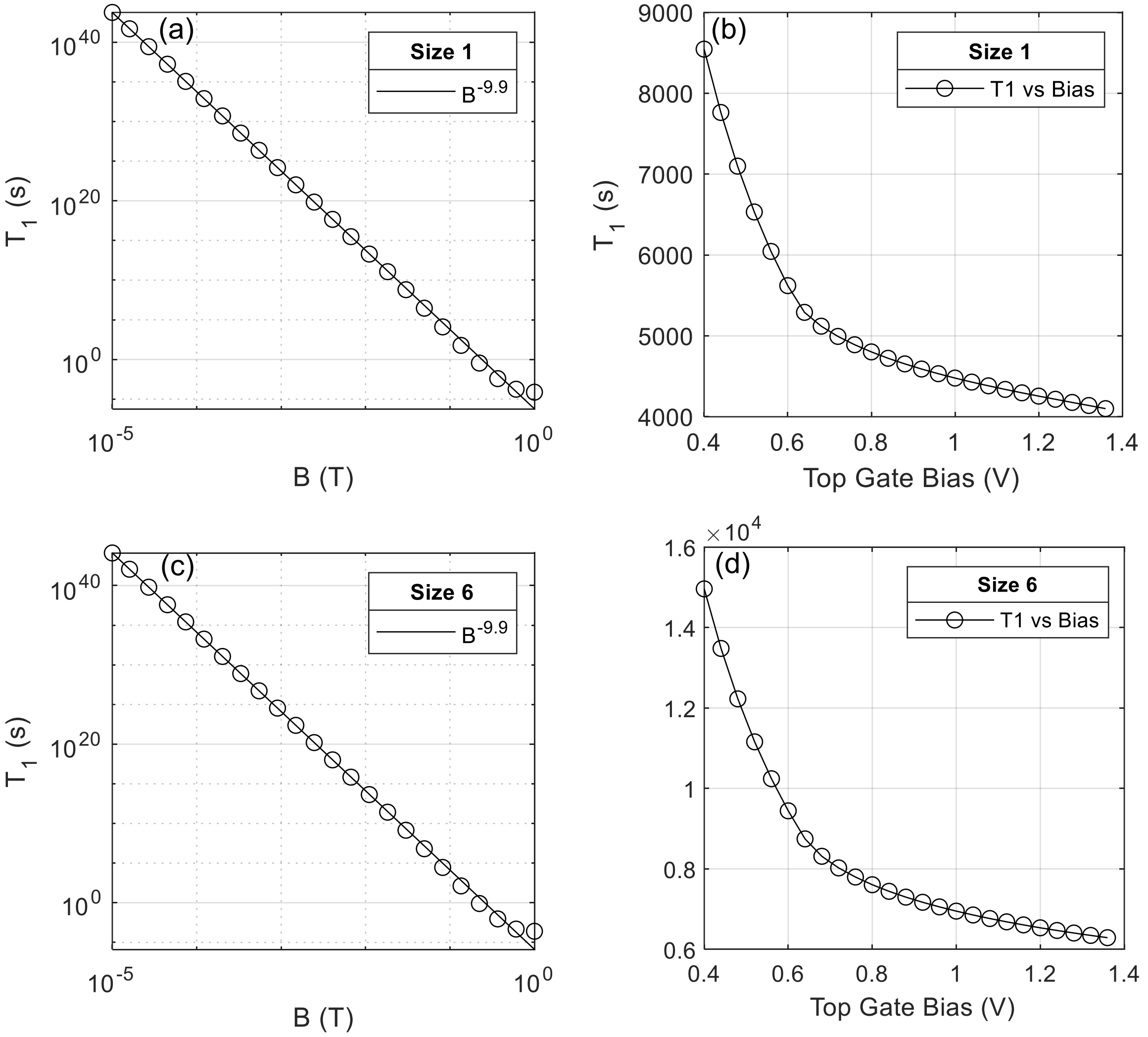

Figure 12: Calculated spin relaxation time $T_1$as a function of out-of-plane magnetic field strength for a fixed gate-bias configuration of 0.6/0.8 V: (a) Size 1 device and (c) Size 6 device. Panels (b) and (d) show the corresponding relaxation time as a function of top-gate bias at a fixed magnetic field of 0.1 T for the Size 1

and Size 6 devices, respectively. The relaxation time is computed using a phonon-mediated spin–orbit interaction model based on Rashba spin–orbit coupling in heavy-hole systems. The dominant relaxation mechanism arises from coupling between confined hole spins and acoustic phonons, enabled by electric-field-induced HH–LH mixing. The Zeeman frequency $\omega_z$, lateral confinement frequency $\Omega$, and HH–LH energy separation, all of which depend on device geometry, enter strongly into the relaxation rate expression.

The Zeeman frequency is given by $\omega_z = g_z \mu_B B/\hbar$, and therefore depends directly on the out-of-plane effective g-factor $g_z$, which was shown in Section III.IV to vary strongly with geometry. As device size increases, $g_z$ increases, which enhances $\omega_z$ and tends to increase the relaxation rate. At the same time, $\Omega$ and $\omega_\pm$ depend strongly on the lateral confinement potential, which is set primarily by the gate-defined electrostatic landscape rather than by the magnetic confinement frequency $\omega_c$. Since the characteristic confinement length satisfies $l_0 = \sqrt{\hbar/(m\omega_0)}$, stronger effective lateral confinement corresponds to larger $\omega_0$ and smaller $l_0$. In the present structures, the size dependence of the confinement potential modifies $\Omega$ in a way that reduces the overall relaxation rate for larger devices, as reflected in Figure 12. The resulting $T_1$ therefore emerges from a competition between the size dependence of $g_z$, the confinement frequencies, and the HH-LH admixture.

Asymmetric gate-bias control affects the relaxation rate through several mechanisms. First, the Zeeman frequency is modulated through the bias dependence of the effective g-tensor discussed in previous sections. Second, under asymmetric bias control, both the position and the extent of the wavefunction change (Section III.III), which alters the wavefunction spread $F$ and therefore the phonon-coupling integral. Third, the gate bias modifies the lateral confinement potential directly, leading to changes in $\omega_\pm$ and $\Omega$. Finally, the Rashba coefficient $\alpha$ itself is electric-field dependent, so bias modulation directly affects the SOC strength entering Eq. $(14)$. As a consequence, the simulated relaxation time exhibits a strong dependence on both device geometry and gate bias.

To assess the impact of out-of-plane magnetic field strength, we compute $T_1 = 1/W_{1n}^R$ as a function of magnetic field for each device size under a fixed gate bias of 0.64/0.84 V. The results are shown in Figure 12 on a logarithmic scale. In addition to g-factor and confinement effects, we account for HH–LH subband coupling by computing the HH–LH splitting from a dedicated numerical procedure based on a one-dimensional reduction of the Luttinger–Kohn Hamiltonian (Appendix D). Specifically, we define the HH–LH energy gap as the energy difference between the heavy-hole ground state and the lowest-energy light-hole state obtained from the same heterostructure potential. This approach ensures consistent extraction of the subband splitting under varying electric field, magnetic field, and strain conditions. From calculations involving up to the 1000$^{\text{th}}$ excited state, we extract HH–LH gaps of 61.9 meV (Size 1) and 66.4 meV (Size 6) significantly smaller than the commonly assumed $\Delta \sim 100$ meV used in simplified models for Ge quantum dots.

To further clarify the role of geometry, Figure 13 shows the relaxation time $T_1$ as a function of device dimension for two representative bias points, 0.40 V and 0.64 V. In both cases, $T_1$ increases nearly linearly with device size, indicating that the larger devices in this structure are less susceptible to phonon-mediated spin relaxation. At the same time, the separation between the two curves demonstrates a distinctly nonlinear bias dependence. This behavior follows naturally from the fact that gate bias modifies not only the electrostatic confinement, but also the effective Rashba coupling, the g-tensor, the wavefunction spread, and the HH-LH admixture. The trends in Figure 13 therefore reinforce the conclusion that spin relaxation in these Ge hole quantum dots is governed by a coupled dependence on device geometry and bias-induced reshaping of the confinement potential.

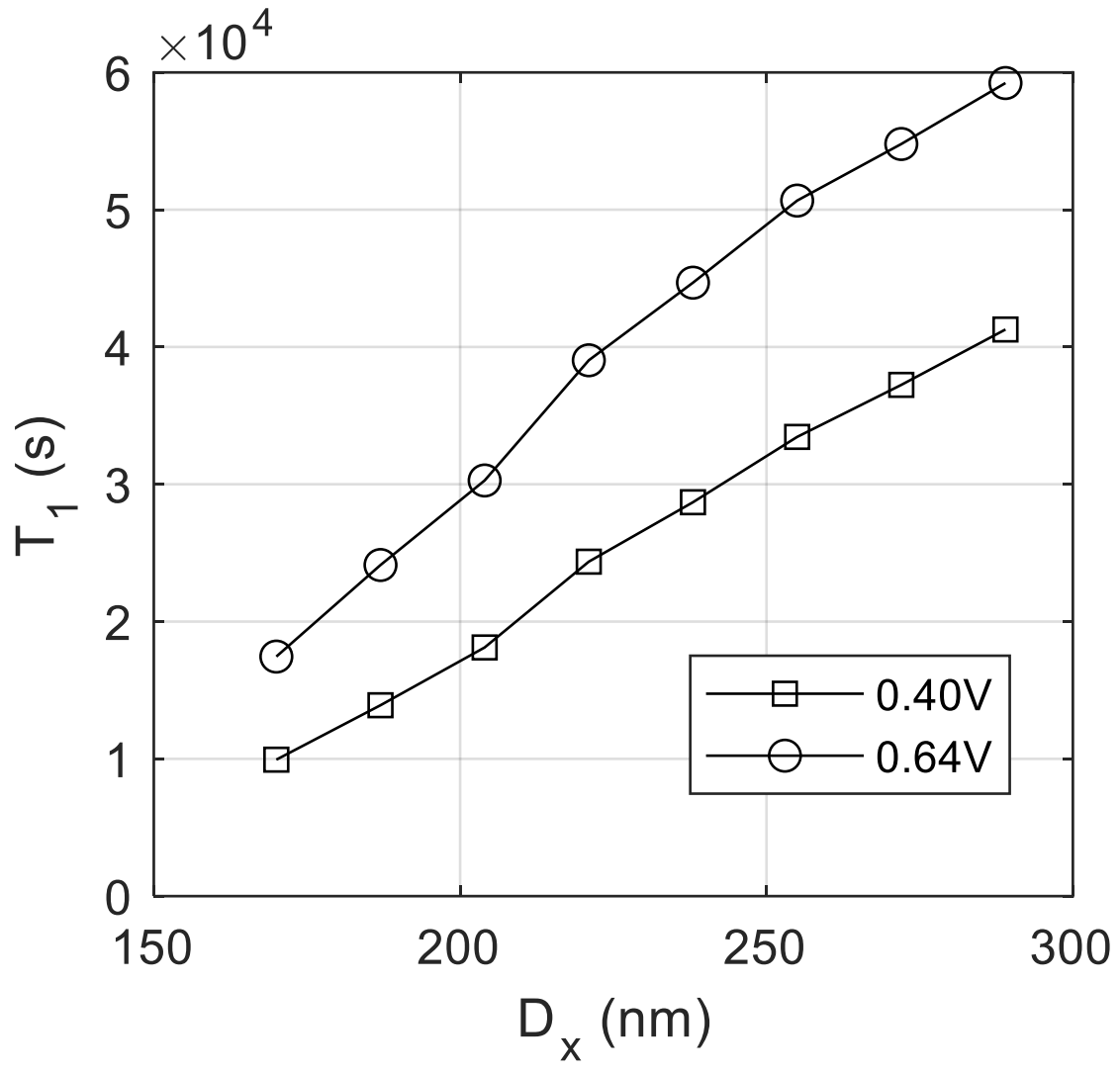

Figure 13: Spin relaxation time $T_1$ as a function of device dimension for simulations performed at gate-bias values of 0.40 V and 0.64 V. At both bias points, $T_1$ increases approximately linearly with device size. The difference between the two curves reflects the nonlinear dependence of spin relaxation on gate bias through its effect on confinement, HH-LH mixing, and spin-orbit coupling.

As expected from this multi-parameter dependence, Figure 12 and Figure 13 show several distinct relaxation regimes in the $T_1(B)$ and $T_1$ (bias) relations. Smaller dots exhibit shorter $T_1$ at fixed magnetic field and bias, consistent with the stronger SOC effects and modified confinement landscape discussed above. In our realistic 3D simulations, the spin-relaxation time shows a steep magnetic-field dependence, scaling approximately as $B^{-9.9}$ for the Size 1 device and $B^{-9.9}$ for the Size 6 device in the low-field regime [Figure 12(a,c)]. This behavior is consistent with theoretical expectations for phonon-mediated relaxation in HH systems dominated by Rashba SOC. Bulaev et al. reported a similar $B^{-9}$ dependence in strongly confined HH dots using a Luttinger model with cubic-in-momentum Rashba terms.[2] Likewise, Woods et al.[22] found $B^{-7}$ to $B^{-9}$ scaling depending on dot geometry and HH-LH splitting in cylindrical HH dots, and Trif *et al.*[23] showed that the dominant one-phonon regime under perpendicular magnetic field also follows a $B^{-7}$-$B^{-9}$ law. Our results therefore agree with the expected $T_1 \propto B^{-n}$ dependence associated with Rashba-mediated spin-phonon coupling in hole systems.

The variation of the fitted exponent with device size [Figure 12(c)] further indicates that lateral confinement and HH-LH mixing modulate the effective Rashba interaction. Under the present asymmetric bias-control scheme, the relaxation time also shows a clear dependence on the effective electric field $E_z$, which is consistent with electrical tunability of the Rashba coefficient and therefore of $T_1$. These results demonstrate that a fully three-dimensional electrostatic model is capable of reproducing, and refining, the expected scaling behavior of phonon-induced spin relaxation in realistic Ge quantum-dot devices.

Overall, spin relaxation in Ge hole spin qubits cannot be described as a simple monotonic function of dot size or magnetic field. Instead, it reflects a coupled dependence on confinement geometry, g-tensor anisotropy, phonon coupling, and HH-LH mixing. This underlines the need for device-specific modeling when predicting relaxation dynamics and suggests that electrically tunable confinement provides a practical route for optimizing $T_1$ without sacrificing gate-based control.

### III.X Device-Design Implications and Model Scope

Taken together, the simulations suggest three experimentally actionable design principles. First, the relative spacing and placement of the confinement gates control the abruptness of wavefunction transfer between the semicircular and approximately rectangular confinement regions and therefore influence the position and width of the g-factor extremum. Second, an additional accumulation gate can shift this extremum to higher bias and broaden the range of reduced first-order $E_z$ sensitivity, providing a means of placing a favorable operating point within a bias window compatible with stable confinement and tunnel-coupling requirements. Third, lateral device scaling modifies $E_{z,\mathrm{eff}}$, HH–LH admixture, g-tensor anisotropy, and $T_1$ even when the vertical heterostructure is unchanged. The dot footprint, gate layout, operating bias, and magnetic-field orientation should therefore be optimized together rather than independently. These correlations arise from the resolved three-dimensional gate potential and are not captured by models that impose a fixed, separable parabolic confinement.

The present calculations remain a continuum-level description. The four-band Luttinger-Kohn model captures the leading HH-LH physics and includes nominal heterostructure strain through the Bir-Pikus Hamiltonian, but it does not explicitly include atomistic alloy disorder, interface roughness, remote-band corrections, or spatially nonuniform strain caused by gate stress and thermal contraction. These effects can alter the absolute g-factors and shift the precise locations of the calculated operating points. The results should therefore be interpreted as robust device-level trends and design principles associated with gate-driven wavefunction redistribution, rather than as atomistically exact predictions for a particular fabricated sample.

## IV. Conclusion

We have shown that spin dynamics in gate-defined Ge hole quantum dots can be strongly engineered through gate geometry and bias asymmetry. In the strained $Si_{0.2}Ge_{0.8}$/Ge/$Si_{0.2}Ge_{0.8}$ devices studied here, the asymmetric gate pattern creates two distinct confinement regimes, and transitions between them drive pronounced changes in wavefunction position, localization, and heavy-hole/light-hole admixture. These changes, in turn, produce strong modulation of the g-tensor and enable geometry-induced dephasing sweet spots, where the qubit becomes first-order insensitive to vertical electric-field fluctuations. This is particularly relevant experimentally, since the bias conditions required to access sweet spots may not always be compatible with other device constraints, making gate geometry an important additional control parameter for engineering usable operating points.

These results were obtained using a high-fidelity, fully three-dimensional simulation framework that combines realistic device electrostatics with a four-band Luttinger–Kohn Hamiltonian. In contrast to simplified models based on idealized two-dimensional geometries or assumed parabolic confinement, the present approach captures the full gate-defined potential landscape, interface-induced asymmetry, and wavefunction redistribution that govern spin-orbit-driven qubit behavior in realistic devices. A central insight of this work is that realistic gate-defined electrostatics do not merely perturb the confined state, but can qualitatively reshape the confinement regime itself and thereby alter the qubit's magnetic response and coherence properties.

We further showed that phonon-induced spin relaxation in these structures depends sensitively on magnetic field, device size, and gate bias. The calculated $T_1$ exhibits a steep magnetic-field dependence close to $B^{-9}$, consistent with Rashba-dominated heavy-hole spin relaxation, while also reflecting the coupled influence of confinement, g-factor anisotropy, and HH–LH mixing. This demonstrates that relaxation in Ge hole qubits cannot be understood from magnetic field or dot size alone, but must be treated as a device-specific property of the full electrostatic and band-structure environment.

The present model remains a continuum-level description. The four-band Luttinger–Kohn Hamiltonian includes the nominal heterostructure strain through the Bir–Pikus formalism, but it does not explicitly account for remote-band corrections, atomistic alloy disorder, interface roughness, or spatially nonuniform strain arising from gate stress and thermal contraction. These effects may alter the absolute g-factors, HH–LH splittings, and precise locations of the calculated operating points. The central conclusions of this work, however, concern gate-driven changes in the resolved three-dimensional confinement potential and the correlated trends among wavefunction redistribution, g-tensor modulation, dephasing sensitivity, and spin relaxation. The present framework is therefore appropriate for identifying device-level trends and design principles, while quantitative comparison with a particular fabricated device would require the additional atomistic and disorder effects noted above.

Overall, this work establishes gate geometry and electrostatic biasing as practical design parameters for tuning g-factor anisotropy, dephasing response, and relaxation dynamics in Ge hole spin qubits. More broadly, it shows that fully three-dimensional, device-resolved modeling is essential not only for quantitative accuracy, but also for identifying confinement-regime transitions and engineered operating points that would be missed in simplified models. These results provide concrete guidance for the design of scalable, electrically controlled quantum-dot qubit architectures in group-IV semiconductors.

## V. Acknowledgment

This study was supported by AFOSR and the Laboratory for Physical Sciences (LPS) under contract numbers FA9550-23-1-0302 and FA9550-23-1-0763.

## VI. Appendix

### A. Luttinger – Kohn – Foreman model

The top four valence subbands (two heavy-hole and two light-hole states) are described using the Luttinger–Kohn–Foreman model, which modifies the standard Luttinger–Kohn Hamiltonian to account for abrupt material interfaces in heterostructures.[24] The basis states are total angular momentum eigenstates |j = 3/2, m⟩, expressed in terms of Bloch functions and spinors:

$$\left[\frac{3}{2},+\frac{3}{2}\right\rangle = \frac{1}{\sqrt{2}}[(X+iY)\uparrow\rangle$$

$$\left[\frac{3}{2},-\frac{3}{2}\right\rangle = \frac{1}{\sqrt{2}}[(X-iY)\downarrow\rangle$$

$$\left[\frac{3}{2},+\frac{1}{2}\right\rangle = \frac{1}{\sqrt{6}}[(X+iY)\downarrow\rangle - \sqrt{\frac{2}{3}}[Z\uparrow\rangle$$

$$\left[\frac{3}{2},-\frac{1}{2}\right\rangle = \frac{-1}{\sqrt{6}}[(X-iY)\uparrow\rangle - \sqrt{\frac{2}{3}}[Z\downarrow\rangle$$

The effective Schrödinger equation is:

$$\begin{pmatrix} P+Q & 0 & -S_- & R \\ 0 & P+Q & -R^\dagger & -S_+ \\ -S_-^\dagger & -R & P-Q & C \\ R^\dagger & -S_+^\dagger & C^\dagger & P-Q \end{pmatrix} \begin{pmatrix} F_{\frac{3}{2},\frac{3}{2}} \\ F_{\frac{3}{2},-\frac{3}{2}} \\ F_{\frac{3}{2},\frac{1}{2}} \\ F_{\frac{3}{2},-\frac{1}{2}} \end{pmatrix} = E \begin{pmatrix} F_{\frac{3}{2},\frac{3}{2}} \\ F_{\frac{3}{2},-\frac{3}{2}} \\ F_{\frac{3}{2},\frac{1}{2}} \\ F_{\frac{3}{2},-\frac{1}{2}} \end{pmatrix}$$

Where

$$P = E_V(r) + \frac{\hbar^2}{2m_e}\left(k_x\gamma_1k_x + k_y\gamma_1k_y + k_z\gamma_1k_z\right)$$

$$Q = \frac{\hbar^2}{2m_e}\left(k_x\gamma_2k_x + k_y\gamma_2k_y - 2k_z\gamma_2k_z\right)$$

$$R = -\frac{\hbar^2\sqrt{3}}{2m_e}k_-\bar{\gamma}k_- + \frac{\hbar^2\sqrt{3}}{2m_e}k_+\bar{\gamma}k_+$$

$$S_\pm = -\frac{\hbar^2\sqrt{3}}{2m_e}[k_\pm(\sigma-\delta)k_z + k_z\pi k_\pm]$$

$$C = \frac{\hbar^2}{m_e}[k_z(\sigma-\delta-\pi)k_- - k_-(\sigma-\delta-\pi)k_z]$$

where:

$k_\pm = k_x \pm ik_y$, $k_\parallel^2 = k_x^2 + k_y^2$, $\bar{\gamma} = \frac{1}{2}(\gamma_3+\gamma_2)$, $\mu = \frac{1}{2}(\gamma_3-\gamma_2)$,
$\sigma = \bar{\gamma} - \frac{1}{2}\delta$, $\pi = \mu + \frac{3}{2}\delta$, and
$\delta = \frac{1}{9}(1+\gamma_1+\gamma_2-3\gamma_3)$

## B. Strain: Bir – Pikus model

Strain effects on the valence band states are modeled using the 4×4 Bir–Pikus Hamiltonian:

$$H_\varepsilon = \begin{pmatrix} P_\varepsilon + Q_\varepsilon & 0 & -S_\varepsilon & R_\varepsilon \\ 0 & P_\varepsilon + Q_\varepsilon & -R_\varepsilon^* & -S_\varepsilon^* \\ -S_\varepsilon^* & R_\varepsilon & P_\varepsilon - Q_\varepsilon & 0 \\ R_\varepsilon^* & -S_\varepsilon & 0 & P_\varepsilon - Q_\varepsilon \end{pmatrix}$$

with:

$$P_\varepsilon = -a_v(\varepsilon_{xx} + \varepsilon_{yy} + \varepsilon_{zz})$$

$$Q_\varepsilon = \frac{-b}{2}(\varepsilon_{xx} + \varepsilon_{yy} - 2\varepsilon_{zz})$$

$$R_\varepsilon = \frac{\sqrt{3}}{2} b(\varepsilon_{xx} - \varepsilon_{yy}) - id\varepsilon_{xy}$$

$$S_\varepsilon = -d(\varepsilon_{xz} - i\varepsilon_{yz})$$

The strain tensor components $\varepsilon_{ij}$ are computed from the symmetrized displacement gradient:

$\varepsilon_{ij} = \frac{1}{2}\left(\frac{\partial u_i}{\partial x_j} + \frac{\partial u_j}{\partial x_i}\right)$ with **u**(x, y, z) as the strained induced displacement field.

## C. Magnetic field effect

*Zeeman effect:*

The Zeeman Hamiltonian includes both linear and cubic terms in total angular momentum $J$:

$$\mathrm{H_z} = -2\frac{\mu_B \kappa}{\hbar}\boldsymbol{J} \cdot \mathbf{B} - 2\frac{\mu_B q}{\hbar}\boldsymbol{\mathcal{J}} \cdot \mathbf{B}$$

Where $\boldsymbol{J} = (J_x, J_y, J_z)$ and $\boldsymbol{\mathcal{J}} = (J_x^3, J_y^3, J_z^3)$, with standard values for Ge: κ = 3.41 and q = 0.06.

*Orbital effect:*

The orbital interaction is included via minimal coupling, where:

$$\boldsymbol{k} \rightarrow \boldsymbol{k} - \frac{Q\boldsymbol{A}}{\hbar}$$

Under the symmetric gauge, the vector potential is:

$$\boldsymbol{A} = \frac{\boldsymbol{B} \times (r - r_0)}{2}$$

## D. Numerical extraction of the HH–LH orbital splitting used to evaluate the Rashba coefficient ($\alpha$)

To quantify the HH–LH splitting used in the spin relaxation calculations, we performed an independent numerical analysis based on a one-dimensional reduction of the Luttinger–Kohn Hamiltonian along the growth direction z. This approach is consistent with the multiband formalism described in Appendix A (Luttinger–Kohn–Foreman model) and Appendix B (Bir–Pikus strain Hamiltonian), but focuses on extracting subband energies under varying electric field, magnetic field, and strain conditions.

### D1. Effective 1D Hamiltonian

Assuming translational invariance in the in-plane directions ($k_x = k_y = 0$), the full multiband Hamiltonian reduces to decoupled HH and LH blocks. The resulting effective Hamiltonians take the form:

$$H_{\text{HH}} = T_{\text{HH}} + V_{\text{HH}}(z), \qquad H_{\text{LH}} = T_{\text{LH}} + V_{\text{LH}}(z)$$

where the kinetic terms are given by:

$$T_{\text{HH}} = \frac{\hbar^2}{2m_0}(\gamma_1 - 2\gamma_2)k_z^2, \qquad T_{\text{LH}} = \frac{\hbar^2}{2m_0}(\gamma_1 + 2\gamma_2)k_z^2$$

and the effective potentials include band offsets, electric field, and strain contributions:

$$V_{\text{HH}}(z) = U(z) + P_\varepsilon + Q_\varepsilon, V_{\text{LH}}(z) = U(z) + P_\varepsilon - Q_\varepsilon$$

Here, $U(z)$ includes the valence band profile and electric field term:

$$U(z) = -E_v(z) + eE_z z$$

and $P_\varepsilon, Q_\varepsilon$ are the strain terms defined in Appendix B.

An out-of-plane magnetic field $B_z$ is included through the Zeeman Hamiltonian:

$$\Delta E = 2\mu_B B_z\left(\kappa m_j + q m_j^3\right)$$

where $m_j = \pm 3/2$ for HH states and $m_j = \pm 1/2$ for LH states. This leads to spin-split HH and LH subbands, from which the lowest-energy branch is selected for each band.

### D2. Numerical Solution and State Tracking

The resulting Hamiltonians are discretized along the $z$-direction using a finite-difference scheme and solved using sparse eigensolvers. For each parameter sweep (electric field $E_z$, magnetic field $B_z$, or strain $\varepsilon_{xx}$), multiple eigenstates are computed.

To ensure consistent tracking of the HH and LH ground states across parameter sweeps, we employ a localization- and overlap-based selection procedure:

- Eigenstates are filtered based on their spatial localization within the QW region.
- A tracking algorithm selects the state that maximizes overlap with the previously identified eigenstate.
- A combined scoring function is used to ensure continuity:

$$\text{score} = w_{\text{ov}} \cdot \langle \psi_{n-1} \mid \psi_n \rangle + w_{\text{loc}} \cdot P_{\text{loc}} - w_E \cdot |E_n - E_{n-1}|$$

where $P_{\text{loc}}$ is the probability density within the well region.

This approach avoids branch switching and ensures that the same physical state is followed across the parameter space.

Next, the HH–LH splitting is defined as:

$$\Delta_{\mathrm{orb}} = E_{\mathrm{LH}} - E_{\mathrm{HH}}$$

where $E_{\mathrm{HH}}$ and $E_{\mathrm{LH}}$ correspond to the lowest-energy heavy-hole and light-hole states, respectively. The quantity extracted here is used as the orbital splitting $\Delta_{\mathrm{orb}}$ entering Eq. (7). Although Eq. (9) provides a simple analytical approximation, the relaxation and Rashba-coefficient plots in the main text use the numerically evaluated $\Delta_{\mathrm{orb}}$ obtained from the 1D Luttinger–Kohn calculation.

This quantity is evaluated as a function of vertical electric field $E_z$, magnetic field $B_z$, and strain $\varepsilon_{xx}$.

The extracted HH–LH splitting serves as a key input to the phonon-induced spin relaxation model discussed in Section III.VII. In particular, $\Delta_{\mathrm{orb}}$ determines the strength of HH–LH mixing, which directly influences the Rashba spin–orbit coupling and the resulting relaxation rate $T_1^{-1}$.

Unlike simplified models that assume a fixed splitting (typically $\sim 100$ meV), our approach reveals that $\Delta_{\mathrm{orb}}$ is strongly dependent on device geometry, strain, and electrostatic conditions, and can be significantly smaller in realistic gate-defined quantum dots. This highlights the importance of self-consistent, device-specific modeling when evaluating spin relaxation mechanisms.

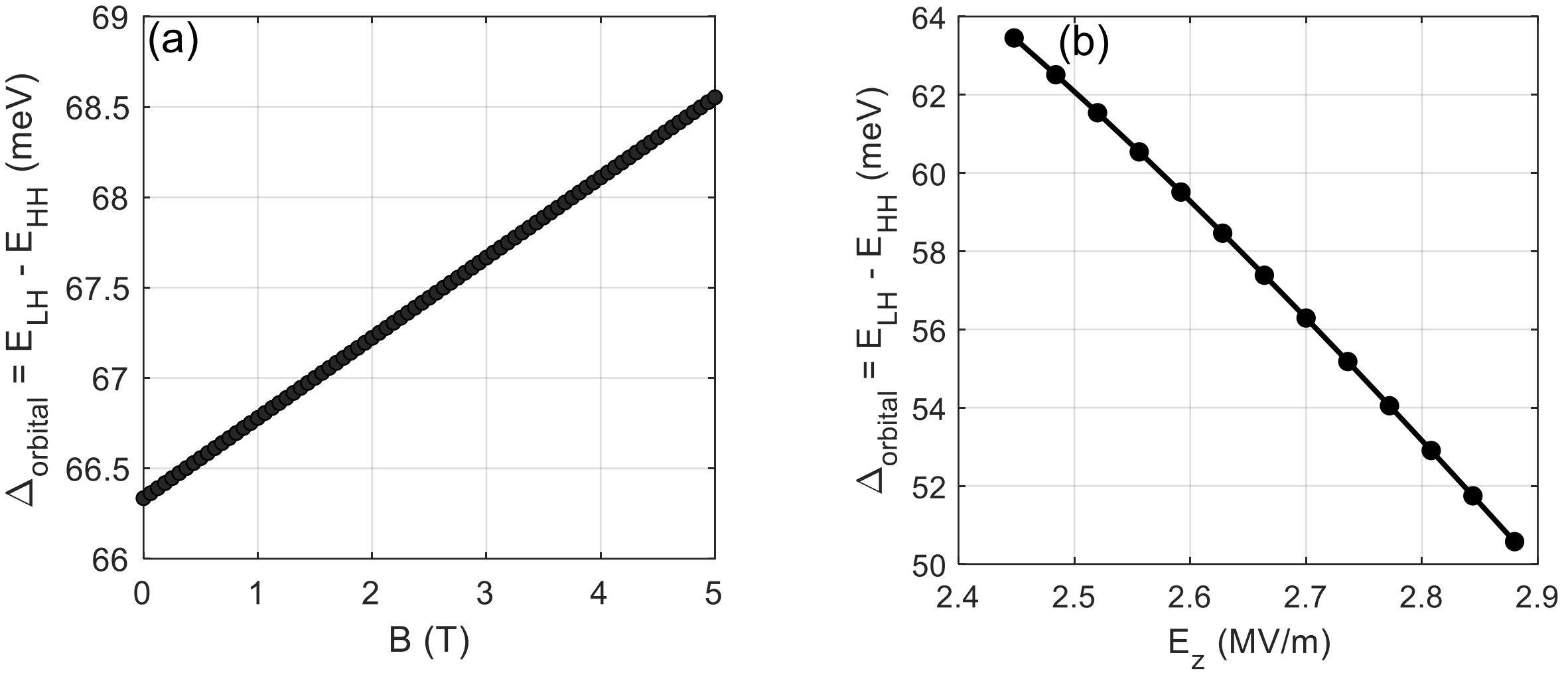

Figure D2: Heavy-hole–light-hole energy splitting $\Delta_{\mathrm{orb}} = E_{\mathrm{LH}} - E_{\mathrm{HH}}$ as a function of (a) vertical electric field $E_z$, (b) strain $\varepsilon_{xx}$, and (c) magnetic field $B_z$, extracted from the one-dimensional Luttinger–Kohn model. The splitting is obtained by tracking the lowest-energy localized HH and LH states across parameter sweeps. The results show strong dependence of $\Delta_{\mathrm{orb}}$ on electrostatic and material parameters, emphasizing the importance of realistic modeling for spin relaxation calculations.

## E. Spin relaxation from a one-phonon Fermi's-golden-rule model

To benchmark the relaxation trends obtained in the main text, we also calculate the spin-relaxation rate using a one-phonon Fermi's-golden-rule approach within the four-band Luttinger–Kohn model. This calculation is adapted from the six-band formalism of Li et al.,[25] but reduced here to the four-band basis used in our simulations. The same material and device parameters are used as in the main text.

Figure D1 shows that the resulting relaxation time $T_1$ has a much weaker magnetic-field dependence than the Rashba-mediated model adopted in Section III.VII. The low-field behavior is closer to the $B^{-4}$-type scaling reported by Li et al.,[25] whereas the main-text model gives a significantly steeper dependence. This comparison highlights that the predicted field dependence of $T_1$ is strongly model dependent and reflects the underlying relaxation mechanism included in each treatment. At magnetic fields above about 0.1 T, the Fermi's-golden-rule results become less reliable because of the dipole approximation used in the matrix elements.

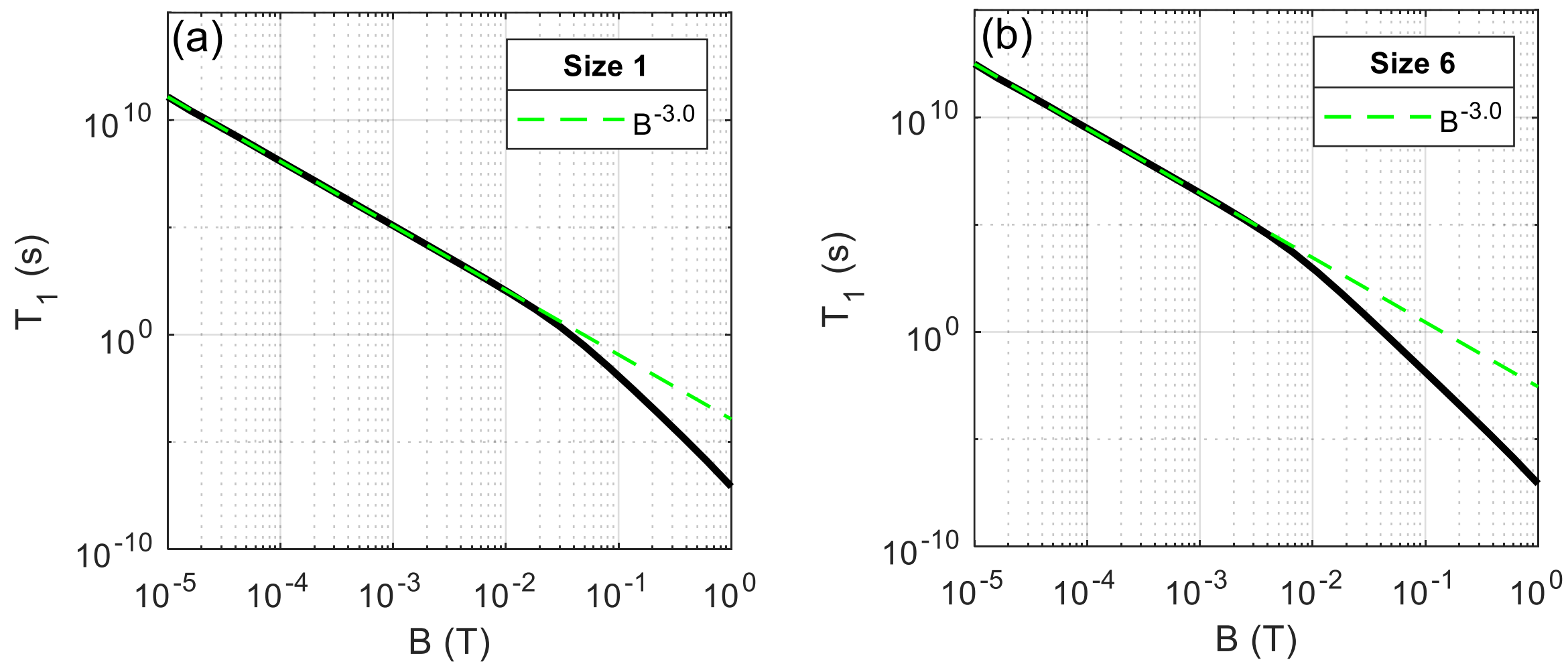

Figure D1: Spin relaxation time $T_1$ as a function of out-of-plane magnetic field, calculated using a one-phonon Fermi's-golden-rule approach within the four-band Luttinger–Kohn model. The magnetic-field dependence is substantially weaker than that obtained from the Rashba-mediated relaxation model used in the main text, and is closer to the low-field $B^{-4}$-type behavior reported by Li et al.[25] Results for $B \gtrsim 0.1$ T have a larger uncertainty because the calculation uses the dipole approximation.

[15] Mauro, Lorenzo, Esteban A. Rodríguez-Mena, Marion Bassi, Vivien Schmitt, and Yann-Michel Niquet. "Geometry of the dephasing sweet spots of spin-orbit qubits." *Physical Review B* 109, no. 15 (2024): 155406.
[16] Hetényi, Bence, Stefano Bosco, and Daniel Loss. "Anomalous zero-field splitting for hole spin qubits in Si and Ge quantum dots." *Physical Review Letters* 129, no. 11 (2022): 116805.
[17] Malkoc, Ognjen, Peter Stano, and Daniel Loss. "Charge-noise-induced dephasing in silicon hole-spin qubits." *Physical Review Letters* 129, no. 24 (2022): 247701.
[18] Yoneda, Jun, Kenta Takeda, Tomohiro Otsuka, Takashi Nakajima, Matthieu R. Delbecq, Giles Allison, Takumu Honda et al. "A quantum-dot spin qubit with coherence limited by charge noise and fidelity higher than 99.9%." *Nature nanotechnology* 13, no. 2 (2018): 102-106.
[19] Volmer, Mats, Tom Struck, Arnau Sala, Bingjie Chen, Max Oberländer, Tobias Offermann, Ran Xue et al. "Mapping of valley splitting by conveyor-mode spin-coherent electron shuttling." *npj Quantum Information* 10, no. 1 (2024): 61.
[20] Moriya, Rai, Kentarou Sawano, Yusuke Hoshi, Satoru Masubuchi, Yasuhiro Shiraki, Andreas Wild, Christian Neumann et al. "Cubic Rashba spin-orbit interaction of a two-dimensional hole gas in a strained-Ge/SiGe quantum well." *Physical review letters* 113, no. 8 (2014): 086601.
[21] Bulaev, Denis V., and Daniel Loss. "Spin relaxation and anticrossing in quantum dots: Rashba versus Dresselhaus spin-orbit coupling." *Physical Review B—Condensed Matter and Materials Physics* 71, no. 20 (2005): 205324.
[22] Woods, L. M., T. L. Reinecke, and R. Kotlyar. "Hole spin relaxation in quantum dots." *Physical Review B* 69, no. 12 (2004): 125330.
[23] Trif, Mircea, Pascal Simon, and Daniel Loss. "Relaxation of hole spins in quantum dots via two-phonon processes." *Physical review letters* 103, no. 10 (2009): 106601.
[24] Foreman, Bradley A. "Effective-mass Hamiltonian and boundary conditions for the valence bands of semiconductor microstructures." *Physical Review B* 48, no. 7 (1993): 4964.
[25] Li, Jing, Benjamin Venitucci, and Yann-Michel Niquet. "Hole-phonon interactions in quantum dots: Effects of phonon confinement and encapsulation materials on spin-orbit qubits." *Physical Review B* 102, no. 7 (2020): 075415.